\documentclass[12pt]{article}
\usepackage[british]{babel}
\makeatletter\AtBeginDocument{\let\@elt\relax}\makeatother
\def\AnswerYes{y}
\def\pdflatex{y}                  
\def\ShowLineNumberVersion{n}     
\def\ShowLabelsVersion{n}         
\def\ShowChangesVersion{n}        
\def\ShowAnnotationsVersion{n}    
\def\ShowFigures{y}               
\def\feynVersion{n}               
\def\MakeArXivLinksActive{y}      
\ifx\pdflatex\AnswerYes
   \usepackage{grffile} 
\fi
\ifx\pdflatex\AnswerYes
    \usepackage{color}
\else
    \usepackage[dvips]{color}
\fi
\ifx\pdflatex\AnswerYes
    \usepackage{thumbpdf}    %
\else
    \usepackage[dvips]{thumbpdf}
\fi
\ifx\pdflatex\AnswerYes
    \usepackage[ 
        final,
        breaklinks=true,
        colorlinks=true,           
        pdfborder={0 0 1},    
        citecolor={blue},linkcolor={blue},urlcolor={red},
        citebordercolor={1 0 0},linkbordercolor={0 0 1},urlbordercolor={1 0 0},
        pdfpagemode=UseOutlines,
        bookmarks=true,bookmarksopenlevel=4
        ]{hyperref}         
\else
    \usepackage[dvips,              
        final,
        breaklinks=true,
        colorlinks=true,      
        pdfborder={0 0 1},    
        citecolor={blue},linkcolor={blue},urlcolor={red},
        citebordercolor={1 0 0},linkbordercolor={0 0 1},urlbordercolor={1 0 0},
        pdfpagemode=UseOutlines,
        bookmarks=true,bookmarksopenlevel=4
         ]{hyperref}         
\fi
\ifx\MakeArXivLinksActive\AnswerYes
   \usepackage{xparse}
   \NewDocumentCommand{\arxiv} %
   {r [: u{ [} u{]]} }{[\href{http://arxiv.org/abs/#2}{arXiv:#2}~[#3]]}
   \NewDocumentCommand{\arxivold} {r[]}{[\href{http://arxiv.org/abs/#1}{#1}]}
   \NewDocumentCommand{\arXiv} %
   {r [: u{ [} u{]]} }{[\href{http://arxiv.org/abs/#2}{arXiv:#2}~[#3]]}
   \NewDocumentCommand{\arXivold} {r[]}{[\href{http://arxiv.org/abs/#1}{#1}]}
\else
   \newcommand{\arxiv}[1][]{[#1]}
   \newcommand{\arxivold}[1][]{[#1]}
   \newcommand{\arXiv}[1][]{[#1]}
   \newcommand{\arXivold}[1][]{[#1]}
\fi
\usepackage{doi} 
\usepackage{orcidlink}
\usepackage[UKenglish]{isodate} 
\usepackage[numbers,sort&compress]{natbib}
\ifx\ShowFigures\AnswerYes
   \usepackage{graphicx}          
\else
   \usepackage[draft]{graphicx}   
   \renewcommand{\includegraphics}[2][]{\fbox{#2}}
\fi
\graphicspath{{figures/}{./}}
\usepackage{multirow}                   
\usepackage{amssymb}
\usepackage{amsmath}
\usepackage{bbm} 
\usepackage{xspace}                    
\xspaceaddexceptions{[]}
\usepackage{dcolumn}                    
\usepackage{pbox}

\usepackage{bm}                        

\usepackage{cancel}	                  

\usepackage{rotating}

\usepackage{floatpag}           
\floatpagestyle{plain}          
\usepackage{slashed}
\ifx\feynVersion\AnswerYes
   \usepackage{feynmp-auto} 
\fi

\usepackage{chngcntr} 
\counterwithin*{equation}{section} 

\usepackage[bottom]{footmisc} 

\ifx\ShowLabelsVersion\AnswerYes
   \usepackage[color]{showkeys} 
   \definecolor{refkey}{gray}{.5}   
   \definecolor{labelkey}{gray}{.5} 
   
\fi
\ifx\ShowAnnotationsVersion\AnswerYes
   \newcommand{\comment}[1]{{\scriptsize\sffamily\bfseries{#1}}}
   \newcommand{\margin}[1]{\marginpar{\scriptsize\sffamily\bfseries{#1}}}
   \newcommand{\drafty}{\textbf{Draft version \today} \hfill}
\else
   \newcommand{\comment}[1]{}
   \newcommand{\margin}[1]{}
   \newcommand{\drafty}{}
\fi
\ifx\ShowLineNumberVersion\AnswerYes
   \usepackage{lineno}
   \linenumbers
\fi
\ifx\ShowChangesVersion\AnswerYes
   \usepackage{ulem}
    
   \newcommand{\delete}[1]{\sout{#1}}            
   \renewcommand{\emph}[1]{\textit{#1}}           
\else

   \newcommand{\sout}[1]{}
   \newcommand{\xout}[1]{}

   \newcommand{\delete}[1]{}
\fi

\newcommand{\disc}{\discretionary{}{}{}}

\newcommand{\cf}{\textit{cf.}\xspace}
\newcommand{\eg}{\textit{e.g.}\xspace}
\newcommand{\ie}{\textit{i.e.}\xspace}
\newcommand{\etal}{\textit{et al.}\xspace}
\newcommand{\etc}{\textit{etc.}\xspace}

\newcommand{\fs}{\scriptstyle} 

\newcommand{\hqq}{\hspace{1em}}
\newcommand{\hqqq}{\hspace{2em}}
\newcommand{\hqm}{\hspace*{-0.25em}}
\newcommand{\hqmm}{\hspace*{-0.5em}}
\newcommand{\hqmmm}{\hspace*{-1.0em}}

\newcommand{\half}{\frac{1}{2}}

\newcommand{\dd}{\mathrm{d}}

\newcommand{\deint}[2]{\dd^{#1}\;\!\! #2\;}

\newcommand{\vectorwithspace}[1]{\vec{#1}\mkern2mu\vphantom{#1}}
\newcommand{\vect}[1]{\vectorwithspace{#1}}

\newcommand{\kv}{\vectorwithspace{k}}

\newcommand{\pv}{\vectorwithspace{p}}
\newcommand{\qv}{\vectorwithspace{q}}

\newcommand{\bra}{\langle}
\newcommand{\ket}{\rangle}

\newcommand{\mpi}{\ensuremath{m_\pi}}

\newcommand{\fpi}{\ensuremath{f_\pi}}
\newcommand{\MeV}{\ensuremath{\mathrm{MeV}}}

\newcommand{\ChiEFT}{$\chi$EFT\xspace}

\newcommand{\NXLO}[1]{N\ensuremath{{}^{#1}}LO\xspace}

\newcommand{\HIGS}{HI$\gamma$S\xspace}
\newcommand{\threeH}{\ensuremath{{}^3}H\xspace}
\newcommand{\threeHe}{\ensuremath{{}^3}He\xspace}
\newcommand{\fourHe}{\ensuremath{{}^4}He\xspace}

\newcommand{\alphae}{\ensuremath{\alpha_{E1}}}
\newcommand{\betam}{\ensuremath{\beta_{M1}}}

\newcommand{\alphaep}{\ensuremath{\alpha_{E1}^{(\mathrm{p})}}}
\newcommand{\betamp}{\ensuremath{\beta_{M1}^{(\mathrm{p})}}}

\newcommand{\alphaen}{\ensuremath{\alpha_{E1}^{(\mathrm{n})}}}
\newcommand{\betamn}{\ensuremath{\beta_{M1}^{(\mathrm{n})}}}

\newcommand{\alphaes}{\ensuremath{\alpha_{E1}^{(\mathrm{s})}}}
\newcommand{\betams}{\ensuremath{\beta_{M1}^{(\mathrm{s})}}}

\newcommand{\piN}{\pi\mathrm{N}}

\newcommand{\N}{\mathrm{N}}
\newcommand{\p}{\mathrm{p}}
\newcommand{\n}{\mathrm{n}}

\newcommand{\MN}{\ensuremath{M_\mathrm{N}}} 
\newcommand{\DeltaM}{\ensuremath{\Delta_{\scriptscriptstyle M}}} 

\newcommand{\omegalab}{\ensuremath{\omega_\mathrm{lab}}}

\newcommand{\omegacm}{\ensuremath{\omega_\mathrm{cm}}}
\newcommand{\thetalab}{\ensuremath{\theta_\mathrm{lab}}}
\newcommand{\thetacm}{\ensuremath{\theta_\mathrm{cm}}}

\newcommand{\Lambdachi}{\overline{\Lambda}_\chi}

\newcommand{\calO}{\mathcal{O}}

\newcommand{\mytitle}[1]{\begin{center}\LARGE{\textbf{#1}}\end{center}}
\newcommand{\myauthor}[1]{\textbf{#1}}
\newcommand{\myaddress}[1]{\textit{#1}}
\newcommand{\mypreprint}[1]{\begin{flushright}#1\end{flushright}}

\newcommand{\wf}{}
\newcommand{\wfbra}{}

\newcommand{\sep}{}

\newcommand{\jrel}{\ensuremath{j_{12}}}

\usepackage{xstring}
\newcommand{\CG}[6]{\langle {#1} {#2}
  {\IfSubStr{#4}{+}{(#4)}{\IfSubStr{#4}{-}{(#4)}{#4}}} 
  {\IfSubStr{#5}{+}{(#5)}{\IfSubStr{#5}{-}{(#5)}{#5}}}  | 
  {#3} {\IfSubStr{#6}{+}{(#6)}{\IfSubStr{#6}{-}{(#6)}{#6}}} \rangle}

\newcommand{\chiSMSfour}{$\chi$SMSN$^4$LO+$400\MeV$+N$^2$LO3NI\xspace}
\newcommand{\chiSMSfourfive}{$\chi$SMSN$^4$LO+$450\MeV$+N$^2$LO3NI\xspace}

\newcommand{\chiSMSfivefive}{$\chi$SMSN$^4$LO+$550\MeV$+N$^2$LO3NI\xspace}
\begin{document}
%

\begin{titlepage}
  \setcounter{page}{0} \pagenumbering{roman}\mypreprint{
    \drafty
    31st January 2024 \\
    Revised version 8th May 2024\\
  }
  
  
  \mytitle{Compton Scattering on \fourHe with Nuclear One- and
    Two-Body Densities}


\begin{center}
  \myauthor{Harald W.\ Grie\3hammer\orcidlink{0000-0002-9953-6512}$^{abc}$}\footnote{Email:
    hgrie@gwu.edu; permanent address: \emph{a}}, 
  \myauthor{Junjie Liao$^a$},
  \myauthor{Judith
    A.~McGovern\orcidlink{0000-0001-8364-1724}$^{d}$}\footnote{Email: judith.mcgovern@manchester.ac.uk},  
    \\[0.5ex]
    \myauthor{Andreas Nogga\orcidlink{0000-0003-2156-748X}$^{e}$}\footnote{Email: a.nogga@fz-juelich.de}
  \emph{and} 
  \myauthor{Daniel R.~Phillips\orcidlink{0000-0003-1596-9087}$^{fg}$}\footnote{Email: phillid1@ohio.edu; permanent address: \emph{f}}
  
\vspace*{0.2cm}
  \myaddress{$^a$ Institute for Nuclear Studies, Department of Physics, \\The
    George Washington University, Washington DC 20052, USA}
  \\[0.7ex]
  \myaddress{$^b$ Department of Physics, Duke University, Box 90305, Durham NC
    27708, USA}
  \\[0.7ex]
  \myaddress{$^c$ High Intensity Gamma-Ray Source, Triangle Universities
    Nuclear Laboratories,\\ Box 90308, Durham NC 27708, USA}
  \\[0.7ex]
  \myaddress{$^d$ Department of Physics and Astronomy, The University of
    Manchester,\\ Manchester M13 9PL, UK}
  \\[0.7ex]
  \myaddress{$^e$ IAS-4, IKP-3 and JCHP, Forschungszentrum J\"ulich, D-52428
    J\"ulich, Germany}
  \\[0.7ex]
  \myaddress{$^f$ Department of Physics and Astronomy and Institute of Nuclear
    and Particle Physics,\\ Ohio University, Athens OH 45701, USA}
  \\[0.7ex]
  \myaddress{$^g$Department of Physics, Chalmers University of Technology, SE-41296 G\"oteborg, Sweden}
\end{center}


\begin{abstract}
  We present the first \emph{ab initio} calculation of elastic Compton scattering from \fourHe. It is carried out to $\calO(e^2 \delta^3)$ [\NXLO{3}] 
  in the $\delta$ expansion of \ChiEFT.  At this order
 and for this target, the only free parameters are the scalar-isoscalar electric and magnetic dipole polarisabilities
 of the nucleon. Adopting current values for these yields a parameter-free 
 prediction. This compares favourably with the world data from \HIGS, Illinois and Lund for photon energies $50\;\MeV\lesssim\omega\lesssim120\;\MeV$ within our theoretical uncertainties of $\pm10\%$.
 We predict a cross section up to 7 times that for deuterium. 
 As in \threeHe, this emphasises and tests the key role of meson-exchange 
 currents between $\n\p$ pairs in Compton scattering on light nuclei. We assess the sensitivity of the  cross section and beam asymmetry to the nucleon polarisabilities, providing clear guidance to future experiments seeking to further constrain them. The calculation becomes tractable 
 by use of the Transition Density Method. The one- and two-body densities generated from $5$
 chiral potentials and the AV18$+$UIX potential
 are available using the python package provided at \url{https://pypi.org/project/nucdens/}.
\end{abstract}
\vskip-0.3cm

\noindent
\begin{tabular}{rl}
  Suggested Keywords: &\begin{minipage}[t]{11cm}
    Chiral Effective Field Theory,
    proton, neutron and nucleon
    polarisabilities, \fourHe Compton scattering,
    $\Delta(1232)$ resonance, pion-exchange currents, Transition Density Method
                    \end{minipage}
\end{tabular}
\end{titlepage}

\setcounter{footnote}{0}

\newpage

\setcounter{page}{1}\pagenumbering{arabic}

%
\section{Introduction}
\label{sec:introduction}

The elastic scattering of photons with energies between approximately 50 MeV and the first resonance region from  bound few-nucleon systems,  $\gamma A\to\gamma A$, examines two fundamental and connected aspects of the electromagnetic structure of hadrons and nuclei. First, this coherent Compton scattering reaction yields determinations of the neutron electric and magnetic scalar dipole polarisabilities from data. Second, it showcases the marked impact of two-body currents mediated by charged pion exchange. 
Since both these aspects of QCD's spontaneously and dynamically broken chiral symmetry are leading contributions to the nuclear Compton response at photon energies $\omega\sim m_\pi$, this process can provide profound insight---but only if the answers across different nuclei are consistent.


Indeed, Compton scattering reactions on the proton and the lightest nuclei have been a vigorous area of experimental and theoretical activity over the last twenty-five years, with the goal being to deliver experimental systematic uncertainties and theoretical accuracy that will elucidate these aspects of QCD.
The rationale for and goals of this international effort are described in a recent White Paper~\cite{Howell:2020nob} and review~\cite{Griesshammer:2012we}. 
The nucleon's electric and magnetic dipole
polarisabilities $\alphae$ and $\betam$ characterise the extent to
which the nucleon acquires an induced electric and magnetic dipole moment in 
external electromagnetic fields; they therefore determine the induced radiation
dipoles~\cite{Griesshammer:2012we}. Compton
scattering from the lightest nuclei, where the theory of nuclear electromagnetic responses is most under control, is our best opportunity to determine the value of these fundamental neutron structure parameters.

Chiral Effective Field Theory (\ChiEFT) is used to interpret such data. \ChiEFT describes both the structure of nuclear targets and the electromagnetic operators that encode the coupling of photons to the degrees of freedom inside the nucleus. It is a model-independent approach that organises the low-energy dynamics of strongly interacting systems in powers of a small expansion parameter. 
Theory uncertainties associated with truncation of the \ChiEFT expansion can therefore be estimated.

Thus far, the best neutron polarisability values are
inferred from a 2015/18 \ChiEFT fit to deuteron Compton data~\cite{Myers:2014ace, Myers:2015aba, Griesshammer:2015ahu}. The scalar-isoscalar polarisabilities obtained are\footnote{The
statistical errors are anti-correlated since the Baldin Sum
Rule $\alphaes+\betams=14.5\pm0.4$ was used, combining the findings of
refs.~\cite{OlmosdeLeon:2001zn, Levchuk:1999zy}. A more recent analysis for
the proton gives $\alphaep+\betamp=14.0\pm0.2$~\cite{Gryniuk:2015eza}.
In combination with the neutron value of $15.2\pm0.4$~\cite{Levchuk:1999zy}, this
would change the isoscalar value to $14.6\pm0.3$. }
\begin{equation}
    \label{eq:alphabeta}
    \alphaes=11.1\pm0.6_\mathrm{stat}\pm0.2_\mathrm{BSR}\pm0.8_\mathrm{th}
    \;\;,\;\;
    \betams=3.4\mp0.6_\mathrm{stat}\pm0.2_\mathrm{BSR}\pm0.8_\mathrm{th}
\end{equation}
in the canonical units for scalar polarisabilities of $10^{-4}~\mathrm{fm}^3$
employed throughout.  Note that these works also estimated  the theory uncertainty from omitted higher-order
terms in \ChiEFT. And indeed, those uncertainties---together
with the point-to-point errors of the deuteron Compton data---dominate the final uncertainty on the neutron numbers
because they are markedly larger than the corresponding uncertainties in the case of the proton:
\begin{equation}
  \label{eq:protonpols}
  \alphaep=10.65\pm0.35_{\text{stat}}\pm0.2_\text{Baldin}\pm0.3_\text{th}
  \;\;,\;\; 
  \betamp = 3.15\mp0.35_\text{stat}\pm0.2_\text{Baldin}\mp0.3_\text{th}\;\;.
\end{equation}

This leads to a lack of clarity on the sign of the 
proton-neutron polarisability difference, and only limited knowledge of its size:
\begin{equation}
  \label{eq:pndifference}
  \alphaep-\alphaen=[-0.9\pm1.6_\text{tot}]\;\;.
\end{equation}
Meanwhile, lattice-QCD determinations remain challenging; see~\eg~\cite{Griesshammer:2015ahu, Detmold:2019ghl, Alexandru:2019, Bignell:2020xkf, Wilcox:2021rtt, Wang:2023omf}.  More work is clearly needed for a better understanding of the degree to which proton and neutron
polarisabilities differ: data and theory must be improved such that the overall uncertainty in the extraction of 
$\alphaep-\alphaen$ is shrunk to about one third of its present size. This is especially important as such experimental information will check
a finding 
of the Cottingham Sum Rule for 
the electromagnetic self-energy correction to the proton-neutron mass
difference: 
$\alphaep-\alphaen=[-1.7\pm0.4]$~\cite{Gasser:2015dwa};
cf.~\cite{Thomas:2014dxa, Tomalak:2018dho, Walker-Loud:2019qhh,
  Gasser:2020mzy}. \ChiEFT predicts strong
  $\mpi$-dependence of this difference, which may be related
  to anthropic arguments~\cite{Griesshammer:2015ahu}. 

The same scalar-isoscalar polarisability combinations
$\alphaes:=\half[\alphaep+\alphaen]$ \etc are probed in 
\fourHe as in the isoscalar deuteron. But, because \fourHe is a scalar, the analysis is now free of contamination from spin polarisabilities
which play a nontrivial r\^ole in proton
extractions~\cite{Mornacchi:2022cln}. 
A \fourHe target also has several experimental advantages that mean measurements using it can deliver a
more accurate result.
It is inert and thus safe to handle, liquefies at
relatively high temperatures, and its high dissociation energy makes for a
clear and simple differentiation between elastic and inelastic events even
with detectors of low energy resolution.  Moreover, cross sections are larger by a factor of 5-to-7 than those for $\gamma$-deuteron scattering. 

In this work, we extend the proton, deuteron and \threeHe Compton analysis in \ChiEFT to \fourHe.
This is also the first \emph{ab initio} computation (in the sense of ref.~\cite{Ekstrom:2022yea}) of this reaction, \cf~a phenomenological approach employed in ref.~\cite{Sikora:2017rfk}. It shows that the growth of the Compton cross section with nuclear mass number is driven by the charged-meson-exchange currents discussed in the opening paragraph.
We adopt the polarisability
values of eq.~\eqref{eq:alphabeta} as input so that \ChiEFT has 
no free parameters at the order we consider. The resulting parameter-free calculation predicts the 
correct size and shape of the Compton cross section. 

In the four-nucleon system, the \emph{Transition Density Method}~\cite{Griesshammer:2020ufp} has already been applied for dark matter-nucleus scattering~\cite{deVries:2023hin}; we now employ it in Compton kinematics.
Since this method markedly simplifies and accelerates the
calculation, two-body current implementations become tractable for $A>3$.
It does this by separating the Compton process into
an interaction kernel of nucleons which do react with the photons, and a
background density of nucleons which do not, with the quantum numbers of the latter being traced over before the density is folded with Compton-scattering operators.

The resulting description at $\calO(e^2 \delta^3)$ in \ChiEFT may not
suffice to reliably \emph{extract} polarisabilities from \fourHe
data, but it \emph{does} permit reliable investigations of the \emph{sensitivity} of
observables to the scalar-isoscalar dipole polarisabilities. This is useful for current
planning of experiments---as we previously argued for a calculation at the same order in
\threeHe~\cite{Margaryan:2018opu}. Therefore, a subsidiary goal in this article is an exploratory study
of the size of the \fourHe elastic Compton differential cross section and its sensitivity to the nucleon polarisabilities.

To that end, we concentrate on the energy region $50\;\MeV\lesssim\omega\lesssim120\;\MeV$ where
polarisabilities are most likely to be extracted---this was also the domain studied in the first calculations of 
Compton scattering on deuterium and ${}^3$He~\cite{Beane:1999uq, Choudhury:2007bh, Shukla:2008zc, ShuklaPhD}.
In this r\'egime, coherent scattering of photons on \fourHe as a whole is suppressed. At higher energies, that mechanism only serves to reduce the dependence on the $2\N$ and $3\N$ interactions that describe the
bound system; see sect.~\ref{sec:kernels}.  The coherent propagation of the $A$-nucleon system between 
the two photon interactions of the Compton reaction is a key ingredient 
in the generation of the correct Thomson limit for the nuclear Compton amplitude at lower energies
$\omega\to0$. Meanwhile, at higher energies, $\omega\to\mpi$, the pion-production
threshold poses additional experimental and theoretical issues which we do not
address here. 

The article is organised as follows. In sect.~\ref{sec:formalism}, we describe
the theoretical ingredients of our approach, starting with a brief and intuition-focused recapitulation of
the Transition Density Method (sect.~\ref{sec:densitymethod}). Compton
observables (sect.~\ref{sec:observables}) and the photon-nucleon Compton
kernels we already used in the deuteron and \threeHe
are summarised (sect.~\ref{sec:kernels}), as well as details of the generation of the transition densities 
(sect.~\ref{sec:potentials}). Section~\ref{sec:results} is devoted to our
results, starting with an overview and comparison to data. In
sect.~\ref{sec:uncertainties}, we quantify the theory uncertainties of our
approach as generically $\pm10\%$ in the cross section---possibly rising to $\pm12\%$ in back-scattering at the highest energies. The
sensitivity of the cross section to varying the polarisabilities is assessed
in sect.~\ref{sec:polarisabilities}. The beam asymmetry is considered in
sect.~\ref{sec:beamasymmetry}, followed by a comparison of the \fourHe results
to those on the other light nuclei (sect.~\ref{sec:othertargets}). Summary,
conclusions and future work are the topic of sect.~\ref{sec:conclusions}.
A preview of the findings reported here was published in ref.~\cite{MENU2023}.

\section{Formalism}
\label{sec:formalism}

\subsection{The Transition Density Method}
\label{sec:densitymethod}

\begin{figure}[!htbp]
\begin{center}
  \includegraphics[width=0.9\linewidth]{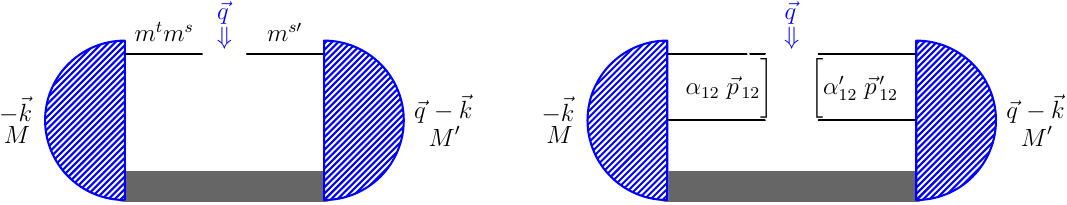}
  \caption{(Colour on-line) Sketches of the one-body (left, $n=1$) and two-body (right, $n=2$)
    transition probability density amplitudes, with pertinent quantum numbers
    and kinematics. The thick grey line denotes the
    $A-n$ spectator nucleons.}
\label{fig:transition}
\end{center}
\end{figure}

We use the \emph{Transition Density Method} as introduced in
refs.~\cite{Griesshammer:2020ufp,deVries:2023hin}.
It factorises the interaction of a probe with a nucleus of $A$
nucleons into an \emph{interaction kernel} between the probe and the $n$
\emph{active nucleons} which directly interact with it, and a backdrop of
$A-n$ \emph{spectator nucleons} which do not. The effect of the spectators is
subsumed into a \emph{$n$-body density}, namely a transition probability
density amplitude that $n$ active nucleons with a specific set of quantum
numbers are found inside the nucleus before the interaction, and are
re-arranged into another specific set of quantum numbers after
it. Figure~\ref{fig:transition} illustrates this separation, with the
interaction kernel depicted as an arrow. It should be stressed that this separation into ``active'' and ``spectator" nucleons is a purely technical one that relies on no approximations beyond those used in more standard evaluations.

An obvious advantage of this factorisation is that densities of the same
nucleus can be recycled with different interaction kernels, while the same
interaction kernels can be used in different nuclei. For example, we checked that the numerical implementation of our \fourHe densities with kernels describing
elastic electron scattering reproduce the well-known electric form factor. Work is also in progress to use the same
\threeHe and \fourHe densities with kernels describing other coherent processes~\cite{Long}. In turn, for Compton scattering on \fourHe, we use the same kernels
as for the \threeHe results~\cite{Margaryan:2018opu}. This brings all Compton
processes under one unified framework. Ref.~\cite{Griesshammer:2020ufp} also demonstrated that results
are numerically identical to that of the more traditional
approach~\cite{Choudhury:2007bh, Shukla:2018rzp, Shukla:2008zc, ShuklaPhD,
  Margaryan:2018opu}, while an orders-of-magnitude reduction of the
computational effort facilitates better-converged numerics. This computational speed-up 
can mainly be attributed to the fact that the production of the densities relies
on well-developed modern numerical few-body techniques, while the kernel
convolutions only involve sums over a limited range of quantum numbers, plus
two three-dimensional integrations in the two-body case; see below. For
\fourHe, production times per energy and angle on 2 nodes with 256 CPUs in total on \textsc{Jureca} are about
$1$ minute (or about $4$ CPU hours) for one-body densities and $3$ minutes (or about $13$ CPU hours) for two-body
ones. Once the densities are in hand, the summation over
one-body quantum numbers is near-instantaneous. For two-body matrix elements,
summation over quantum numbers in the $(12)$ subsystem as well as---since they, too, are undetected---angular and radial integrations over its incoming and outgoing relative momenta (\cf~eq.~\eqref{eq:twobody} below)
adds less than a CPU hour per energy and angle on
a workstation to achieve better-than-$0.7\%$ numerical
accuracy; see sect.~\ref{sec:numerics}.

It is a fundamental advantage of \ChiEFT that it provides a well-defined
procedure to predict a hierarchy of $n$-body interaction
kernels~\cite{Weinberg:1990rz, Weinberg:1991um, Weinberg:1992yk, vanKolckPhD,
  vanKolck:1994yi, Friar:1996zw}; see also refs.~\cite{Epelbaum:2008ga, Phillips:2016mov, Machleidt:2016rvv, Hammer:2019poc, Epelbaum:2019kcf}. We concentrate on one- and two-body kernels in the following since these usually dominate, while three-and-more-body kernels are suppressed in powers of \ChiEFT's small dimensionless expansion parameter. As discussed in sect.~\ref{sec:kernels}, this suppression holds in Compton scattering for $\omega\gtrsim50\;\MeV$ at the order we consider. 

Instead of recounting formal details, we provide now a slightly simplified
account. We leave out some notational subtleties and concentrate on
interactions which neither change the isospin projection of the $n$ active
nuclei, nor explicitly depend on the cm momentum $\kv$ of the probe. The latter
would stem, for example, from boost corrections of the active nucleons. In Compton scattering, both effects are of higher order than considered here.

Specifically, the \emph{one-body density}
$\rho^{M^\prime M}_{m^tm_s^\prime m^s}(\qv)$ ($n=1$, left in
fig.~\ref{fig:transition}) is the transition probability density amplitude to
find nucleon $A$ with specific isospin-spin projection $(m^tm^s)$
inside a nucleus of momentum $-\kv$ and spin projection $M$ in the
centre-of-mass system, to have it absorb a momentum transfer $\qv$ and
re-arrange its spin projection to $m^{s\prime}$, and finally be
re-incorporated into the $A$-body system such that the outgoing nucleus
remains intact and in its ground state---although now with spin projection $M^\prime$. 
The one-body interaction kernel $ O_{A}^{m^tm^{s \prime} m^{s }}(\qv)$ is then characterised by the momentum transfer $\qv$ and the quantum
numbers $(m^tm^{s\prime}m^s)$, and the one-body matrix element is found by
summing over them:
\begin{equation}
  \label{eq:onebody}
  \hqmm\hqm\bra \, \wfbra \sep M ' \, | \, O_{A} \,  | \, \wf
  \sep M \, \ket = 
  \sum\limits_{\genfrac{}{}{0pt}{}{\fs m^{s\prime} m^{s  }}{\fs m^{t}}}
  O_A^{m^t m^{s \prime} m^{s }} ( 
  \qv)\;\rho_{m^tm^{s \prime
    }\,m^{s}}^{M^\prime M}(\qv)\;\;.
\end{equation}
To be specific and remove any ambiguity in its definition, we state that the one-body kernel is the matrix element in the specified kinematics as derived without additional factors from the (non-relativistic) Feynman rules of the pertinent process. For example, in Compton scattering, the LO [$\calO(e^2\delta^0)$] contribution is the one-nucleon Thomson term of fig.~\ref{fig:1Bdiagrams} (a): 
\begin{equation}
    O_{A,\,\mathrm{Thomson}}^{m^t m^{s \prime} m^{s }}=-\frac{(Q^{m^t}e)^2}{M_N}\;\vect{\epsilon}^{\prime\dagger}\cdot\vect{\epsilon}\;\delta^{m^{s \prime} m^{s }}\;\;,
\end{equation} 
with $\vect{\epsilon}$ and $\vect{\epsilon}^\prime$ the polarisation of the incident and outgoing photon, respectively, and $Q^{m^t}=m^t+\half$ the nucleon charge in units of the elementary electric charge $e$.

Likewise, the \emph{two-body density}
$\rho^{M^\prime M}_{\alpha_{12}^\prime\alpha_{12}}(p_{12}^\prime,p_{12};\qv)$
($n=2$, right in fig.~\ref{fig:transition}) is the transition probability
density amplitude for the pair of nucleons $(12)$ with specific quantum numbers $\alpha_{12}$ 
and intrinsic relative momentum $\pv_{12}\equiv p_{12}\;\hat{p}_{12}$ (magnitude $p_{12}$, direction $\hat{p}_{12}$)
to absorb a
momentum transfer $\qv$ resulting in a new relative momentum $\pv_{12}^\prime$
and new quantum numbers $\alpha^\prime_{12}$, before being absorbed back into
the nucleus. Here, $\alpha_{12}$ is a complete set of quantum numbers which
uniquely specifies the pair, including isospin $t_{12}\in\{0;1\}$ and isospin projection
$m^t_{12}$, spin $s_{12}\in\{0;1\}$, total angular momentum $j_{12}$ with its projection
$m_{12}$, and orbital angular
momentum $l_{12}\in\{|j_{12}-s_{12}|;\dots;j_{12}+s_{12}\}$. The two-body kernel
$O_{12}^{\alpha_{12}^\prime m_{12}^{s\prime}\alpha_{12}m_{12}^{s}}(\pv_{12}',
\pv_{12})$ is then characterised by $(\alpha_{12}^\prime
m_{12}^{s\prime}\alpha_{12}m_{12}^s)$ and
relative momenta $(\pv_{12}^\prime, \pv_{12})$. Clebsch-Gordan coefficients $\CG{j_{1}}{j_{2}}{j}{m_1}{m_{2}}{m}$ in the
convention of refs.~\cite{Edmonds, PDG} project its angular dependence in
$\pv_{12}^\prime$ and $\pv_{12}$ onto one on the orbital angular momenta
$(l_{12}^\prime,m^{l\prime}_{12})$ and $(l_{12},m^l_{12})$ of out- and in-states, as constrained by the spin of the nucleon pair and
its total angular momentum. Hence, the additional quantum number 
of the spin-projection $m_{12}^s$ of the pair enters, too. 
A two-body matrix element needs finally
both summation over quantum numbers and integration over the (undetected)
relative momenta of the pair:
\begin{equation}
  \label{eq:twobody}
  \begin{split}
 \bra\wfbra \sep &M ' \, | O_{12} | \wf \sep M \, \ket \equiv\\
  &\sum_{\alpha_{12}'\alpha_{12}} \int 
  \frac{\deint{}{p_{12}'} p_{12}^{\prime \, 2 }}{(2\pi)^3}\;\deint{}{p_{12}} p_{12}^{2} \; 
  \rho_{\alpha_{12}'\,\alpha_{12}}^{
    M'M}
  (p_{12}',p_{12};\qv)\\
  &\hqqq\times\sum_{m^{s \prime }_{12} m^{s}_{12} }  \CG{l_{12}'}{s_{12}'}{\jrel'}
  {m_{12}' - m^{s \prime }_{12}}{m^{s
      \prime}_{12}}{m_{12}'} \CG{l_{12}}{s_{12}}{\jrel}
  {m_{12} - m^{s}_{12}}{m^{s}_{12}}{m_{12}}   \\[0.5ex]
  &\hqqq\hqqq\times \int \deint{}{\hat p_{12}'} \deint{}{\hat p_{12}}
  Y^{\dagger}_{l_{12}' (m_{12}' - m^{s \prime }_{12})} ( \hat p_{12}') \;
  Y_{l_{12} (m_{12} - m^{s }_{12})} ( \hat p_{12})\; O_{12}^{\alpha_{12}'m_{12}^{s\prime} \alpha_{12}m_{12}^{s}} (\pv_{12}^{\,\prime},
  \pv_{12}; \qv ) \;\;.\hspace*{-3ex}
  \end{split}
\end{equation}
In ref.~\cite{Griesshammer:2020ufp}, the integration measure $(2\pi)^{-3}$ had
implicitly been attributed to the kernel but this was not explicitly
stated. Since we now define the kernel as derived without additional factors
from the (non-relativistic) Feynman rules of the pertinent process, this
factor must be included in the convolution. To remove any such ambiguity going
forward, we quote a simple contribution to the \NXLO{2} [$\calO(e^2\delta^2)$]
two-body Compton kernel, namely diagram $\mathrm{(a)}$ in fig.~\ref{fig:2Bdiagrams} as reported first in~\cite{Beane:1999uq} and extended to isospin $t_{12}=1$ in~\cite{Choudhury:2007bh,
  Shukla:2008zc, ShuklaPhD}: 
\begin{eqnarray}
    \label{eq:2Bkernelexample}   
      &&\hqmmm\hqm O_{12,\;\mathrm{diagram\;(a)}}^{\alpha_{12}'m_{12}^{s\prime} \alpha_{12}m_{12}^{s}} (\pv_{12}^{\,\prime},
  \pv_{12}; \kv; \qv )=(-1)^{t_{12}}\;\delta_{m^t_{12}0}\;\frac{g_A^2\;e^2}{2\fpi^2}\\
  &&\hqmm\times\hqmm\sum\limits_{\genfrac{}{}{0pt}{}{\fs m^{s\prime}_1 m^{s}_1}{\fs m^{s\prime}_2 m^s_2}}
  \CG{\half}{\half}{s_{12}^\prime}{m_1^{s\prime}}{m_2^{s\prime}}{m_{12}^{s\prime}}\,
  \CG{\half}{\half}{s_{12}}{m_1^s}{m_2^s}{m_{12}^s}\;
 \frac{\vect{\epsilon}\cdot(\vect{\sigma}_1)^{m^{s\prime}_1}{}_{m^s_1}\;\vect{\epsilon}^{\prime\dagger}\cdot(\vect{\sigma}_2)^{m^{s\prime}_2}{}_{m^s_2}}{\omega^2-\mpi^2-(\vect{p}_{12}-\vect{p}_{12}^\prime+\vect{k}+\frac{\vect{q}}{2})^2}+(1\leftrightarrow2)
 \nonumber \; \;.
\end{eqnarray}
Here, $g_A=1.267$ is the pion-nucleon coupling, $\fpi=92.42\;\MeV$ the pion decay constant, $\mpi=139.5675\;\MeV$ its mass, $(\omega,\kv)$ the incident photon's $4$-momentum, and $(\vect{\sigma}_i)^{m^{s\prime}_i}{}_{m^s_i}$ the components of the Pauli spin matrix of nucleon $i$ which combine to a given $s_{12}$ and $m_{12}^s$. As customary, $(1\leftrightarrow2)$ stands for another term with the r\^oles of the nucleons exchanged, $\vect{p}_{12}\to-\vect{p}_{12}$, $\vect{p}_{12}^\prime\to-\vect{p}_{12}^\prime$, and with $t_{12}+l_{12}+s_{12}$ odd due to the Pauli principle.

At zero momentum transfer, the trace of each one- or two-body density in the space of quantum numbers and momenta is normalised to $1$ (\ie~one nucleon or one nucleon-pair is active); see~\cite{Griesshammer:2020ufp} for details and symmetry properties.

Before closing this section, we note that the probabilistic interpretation of an $n$-body transition density amplitude we laid out here is only rigorously true to the extent that a $n$-body operator can be used to compute the probability amplitude; \cf~some popular and recent accounts~\cite{Polyzou:1997je, More:2017syr, Tropiano:2021qgf}. At momenta of order $\Lambdachi$, dependence on the chiral interactions and on phenomena at momenta above the domain of \ChiEFT enter each $n$-body transition density amplitude. In observables, these are compensated for by other operators involving different $m\ne n$-body densities as well as higher-order effects in both the kernel and potential. For example, high-momentum parts of one-body densities play against those of two-body densities. That makes a simple physical interpretation of the Transition Density along the lines of this subsection inaccurate at such momenta.

\subsection{Compton Observables in \texorpdfstring{\fourHe}{4He}}
\label{sec:observables}

In addition to the quantum numbers of the active nucleons, Compton
scattering is characterised by the quantum numbers of the photons: initial and
final polarisations $\lambda,\lambda^\prime$. The total matrix
element is finally obtained by weighting with the number of indistinguishable
nucleons and nucleon pairs, 
\begin{equation}
  \label{eq:comptonME}
  A_{M\lambda}^{M^\prime\lambda^\prime}(\kv,\qv)=\genfrac{(}{)}{0pt}{}{A}{1}\;\bra \wfbra M^\prime|
  {O}^{\lambda^\prime \lambda}_A|\wf M \ket\;+\;
  \genfrac{(}{)}{0pt}{}{A}{2}\;\bra \wfbra M^\prime|
  {O}^{\lambda^\prime \lambda}_{12}|\wf M \ket\;+\;\dots\;\;,
\end{equation}
 with $A=4$ nucleons inside \fourHe. The ellipses denote more-body kernels which are of higher order, as discussed in sect.~\ref{sec:kernels}.
This amplitude is evaluated in the cm frame of the photon-nucleus system,
where no energy is transferred. In Compton scattering, the 
momentum-transfer $\qv$ which characterises the transition densities
is traditionally replaced by the energy of both
incident and outgoing photon, $\omegacm=|\kv|=|\kv^{\prime}|$ and by the
scattering angle $\thetacm$ for the outgoing photon:
\begin{equation}
\label{eq:momtransfer}
  \cos\thetacm=1-\frac{\qv^{\,2}}{2\omegacm^2} \;\;. 
\end{equation}
The \fourHe Compton cross section (target spin $0$, \ie~$M^\prime=M=0$) in the lab frame is finally found by
transferring angles and energies from the cm using the
\fourHe mass of $3727.4\;\MeV$ and averaging (summing) over initial (final) photon
polarisations:
\begin{equation}
  \label{eq:crosssection}
  \frac{\dd\sigma}{\dd\Omega}=\frac{1}{2}
  \left(\frac{\omegalab^\prime}{4\pi\omegalab}\right)^2\sum\limits_{\lambda^\prime\lambda}
  \left|A_{0\lambda}^{0\lambda^\prime}(\kv,\qv)\right|^2\;\;.
\end{equation}
In (elastic) Compton scattering on a spin-$0$ target like \fourHe, only one
more observable is realistically measurable in today's facilities: the
asymmetry of a linearly polarised beam without measurement of the scattered
photon's polarisation: 
\begin{equation} 
  \label{eq:beamasym}
  \Sigma^\text{lin}\equiv\Sigma_3:=\frac{\dd \sigma^{||}-\dd
    \sigma^{\perp}}{\dd \sigma^{||}+\dd
    \sigma^{\perp}}=-\frac{\sum\limits_{\lambda^\prime\lambda}
    A_{0\lambda}^{0\lambda^\prime}(\kv,\qv)\;
    [A_{0-\lambda}^{0\lambda^\prime}(\kv,\qv)]^\dagger}
  {\sum\limits_{\lambda^\prime\lambda}
  \left|A_{0\lambda}^{0\lambda^\prime}(\kv,\qv)\right|^2}\;\;.
\end{equation}
 Here, $\dd \sigma$ is shorthand for ${\dd\sigma}/{\dd\Omega}$
and superscripts refer to photon polarisations (``$\parallel$'' for
polarisation in the scattering plane, ``$\perp$'' for perpendicular to
it). The kinematic variables must of course be transferred to the lab frame as
appropriate. 

\subsection{Compton Kernel}
\label{sec:kernels}

For reviews of Compton scattering on nucleons and light nuclei in \ChiEFT, we refer the
reader to refs.~\cite{Griesshammer:2012we, McGovern:2012ew} for notation, relevant parts of the chiral Lagrangian, and full
references to the literature. Here, we merely summarise the power counting
and Compton kernel already employed in refs.~\cite{Hildebrandt:2005ix,
  Hildebrandt:2005iw, Margaryan:2018opu} in our region of interest,
$50\;\MeV\lesssim\omega\lesssim120\;\MeV$.

When \ChiEFT with a dynamical Delta is used to compute (elastic) Compton scattering, three low
scales compete: the pion mass $\mpi$, the Delta-nucleon mass splitting
$\DeltaM \approx 300\;\MeV$, and the photon energy $\omega$.  Each provides a
small, dimensionless expansion parameter, measured in units of the ``high'' momentum
scale $\Lambdachi$, at which the theory breaks down because new degrees of
freedom enter\footnote{This physically meaningful parameter is not to be confused with an unphysical ``cutoff" $\Lambda$, albeit the symbols are similar~\cite{Furnstahl:2014xsa,Griesshammer:2021zzz}.}. While $\frac{\mpi}{\Lambdachi}$ and
$\frac{\DeltaM}{\Lambdachi}$ have quite different chiral behaviour, we follow
Pascalutsa and Phillips~\cite{Pascalutsa:2002pi} and take a common breakdown
scale $\Lambdachi \approx 650\;\MeV$, consistent with the masses of the
$\omega$ and $\rho$ as the next-lightest exchange mesons, exploiting a
numerical coincidence at the physical pion mass to define a single expansion parameter:
\begin{equation}
  \delta\equiv\frac{\Delta_M}{\Lambdachi}\approx \sqrt{\frac{m_\pi}{\Lambdachi}}
  \approx\sqrt{\frac{\omega}{\Lambdachi}}\approx 0.4\ll1\;\;.
\end{equation}
We also count $\MN\sim\Lambdachi$.  Since $\delta$ is not very small,
order-by-order convergence must be verified carefully, see sect.~\ref{sec:uncertainties}.

This power counting organises contributions under the assumption $\omega\sim\mpi$.  
As extensively discussed previously~\cite{Beane:1999uq,
  Beane:2004ra,Hildebrandt:2005ix, Hildebrandt:2005iw, Margaryan:2018opu} and
summarised in~\cite[sect.~5.2]{Griesshammer:2012we}, in this r\'egime only kernels with one and
two active nucleons contribute in a \ChiEFT description of Compton scattering up to and including
\NXLO{4} [$\calO(e^2\delta^4)$]. At lower energies $\omega \lesssim m_\pi^2/M$, this power counting does not apply because photons with resolution $1/\omega$ 
larger than the size of the \fourHe nucleus scatter coherently on the whole
target.  Refs.~\cite{Hildebrandt:2005ix, Hildebrandt:2005iw,
  Griesshammer:2012we} discuss in detail how the reformulated power counting appropriate at these lower energies leads to the restoration of
the Thomson limit by inclusion of coherent propagation of the \fourHe system
in the intermediate state between absorption and emission of photons. This
rescattering effect involves the interaction of all $A$ nucleons with one
another between photon absorption and emission, and hence an $A$-body
density. However, that r\'egime is not the focus of this presentation. Rather, we are
concerned with the non-collective contributions which dominate above about
$50\;\MeV$. That is also where data is most likely to be taken to extract
nucleon polarisabilities.

Specifically, we use the one- and two-body Compton kernels of
refs.~\cite{Bernard:1991rq,Bernard:1995dp,Beane:1999uq}, supplemented with the
$\Delta$-pole and $\pi \Delta$ loop graphs~\cite{Hildebrandt:2003fm,
  Hildebrandt:2004hh, Hildebrandt:2005ix, Hildebrandt:2005iw}. They are both
conceptually and numerically identical to the ones which have been described
extensively in our Compton studies of the deuteron~\cite{Hildebrandt:2005ix,
  Hildebrandt:2005iw, Griesshammer:2013vga, Myers:2014ace, Myers:2015aba}
 and, most recently, \threeHe~\cite{Margaryan:2018opu,
  Griesshammer:2020ufp}. These pieces of the photonuclear operator are
organised in a perturbative expansion which is complete up to and including
\NXLO{3} [$\calO(e^2 \delta^3)$]. 
No contribution enters at NLO [$\calO(e^2\delta^1)$], and only one-body Delta contributions at \NXLO{3} [$\calO(e^2\delta^3)$]. We only allow photon energies somewhat below 
$\omega_\mathrm{thr}(\mbox{\fourHe})\approx\mpi$ in order to 
avoid additional complications in the vicinity of the pion-production threshold.

\begin{figure}[!t]
  \begin{center}
    \includegraphics[width=0.8\textwidth]{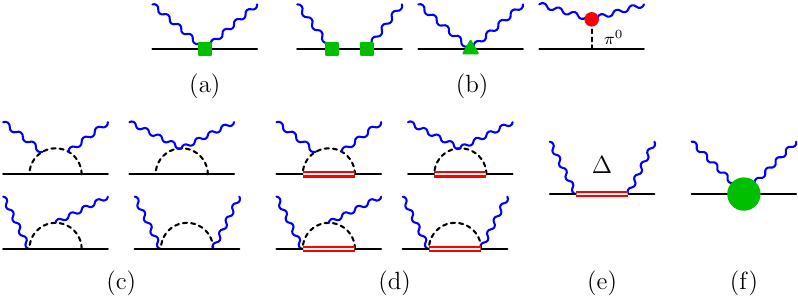}
    \caption{(Colour on-line) The single-nucleon contributions in \ChiEFT up
      to \NXLO{3} [$\calO(e^2\delta^3)$] for
      $50\;\MeV\lesssim\omega\lesssim120\;\MeV$. The vertices are from:
      ${\cal L}_{\piN}^{(1)}$ (no symbol), ${\cal L}_{\piN}^{(2)}$ (green
      square), ${\cal L}_{\piN}^{(3)}$ (green triangle),
      ${\cal L}_{\pi\pi}^{(4)}$ (red disc)~\cite{Bernard:1995dp}; the green
      disc of graph (f) stands for variations of the polarisabilities
      . Permuted and crossed diagrams are not
      displayed.}
    \label{fig:1Bdiagrams}
  \end{center}
\end{figure}

The one-nucleon kernel convoluted with the one-body density as in eq.~\eqref{eq:onebody} is sketched in fig.~\ref{fig:1Bdiagrams}:

\begin{itemize}

\item[(a)] LO [$\calO(e^2\delta^0=Q^2)$]: The single-nucleon (proton) Thomson term.

\item[(b)] \NXLO{2} [$\calO(e^2\delta^2=Q^3)$] non-structure/Born terms:
  photon couplings to the nucleon charge beyond LO, to its magnetic moment, or
  to the $t$-channel exchange of a $\pi^0$ meson (irrelevant for the isoscalar
  \fourHe).

\item[(c)] \NXLO{2} [$\calO(e^2\delta^2=Q^3)$] structure/non-Born terms:
  photon couplings to the pion cloud around the nucleon is the source of the
  LO contributions to the polarisabilities as first reported in
  refs.~\cite{Bernard:1991rq, Bernard:1995dp}.

\item[(d/e)] \NXLO{3} [$\calO(e^2\delta^3)$] structure/non-Born terms: photon
  couplings to the pion cloud around the (non-relativistic) $\Delta(1232)$ (d) or
  directly exciting the Delta (e), as calculated in
  refs.~\cite{Butler:1992ci, Hemmert:1996rw, Hemmert:1997tj}; these give NLO contributions to
  the polarisabilities. The Delta parameters are taken from ref.~\cite{Margaryan:2018opu}. The Delta excitation of diagram (d) shows
  considerable energy dependence even at $\omega\sim\mpi$; see the discussion
  of ``dynamical polarisabilities'' in refs.~\cite{Griesshammer:2012we,
    Griesshammer:2017txw}. This will be important in the interpretation of our results in sect.~\ref{sec:results}.

\item[(f)] Short-distance/low-energy coefficients (LECs) encode those
  contributions to the nucleon polarisabilities which stem from physics at and above the breakdown scale $\Lambdachi$. These offsets to the polarisabilities are formally of higher order. We determine them to reproduce the isoscalar polarisabilities of eq.~\eqref{eq:alphabeta}. Their 
uncertainties were discussed in the Introduction  but are dwarfed by the other uncertainties of the results presented
here, including those coming from the wave function dependence and higher
order effects; see sect.~\ref{sec:uncertainties}. Neither does the detailed discussion of the sources and sizes
of uncertainties of other nucleon polarisabilities in ref.~\cite{Griesshammer:2015ahu}
bear on the present results. We also recall from the Introduction that since \fourHe is a near-perfect isoscalar and
scalar, neither the nucleon's isovector polarisabilities nor its spin
polarisabilities enter, except at very high orders.

\end{itemize}

\begin{figure}[!b]
  \begin{center}
    \includegraphics[width=0.5\textwidth]{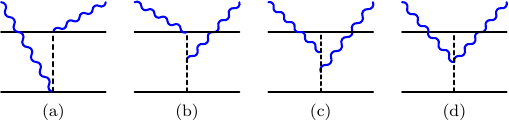}
    \caption{(Colour on-line) \NXLO{2} [$\calO(e^2 \delta^2)$]  contributions to the
      (irreducible) $\gamma \N\N \rightarrow \gamma \N\N$ amplitude (no additional
      contributions at \NXLO{3} [$\calO(e^2 \delta^3)$].  Notation as in
      fig.~\ref{fig:1Bdiagrams}. Permuted and crossed diagrams not displayed.}
    \label{fig:2Bdiagrams}
  \end{center}
\end{figure}

The first nonzero two-body kernel convoluted with the two-body density as in eq.~\eqref{eq:twobody} enters at
\NXLO{2} [$\calO(e^2 \delta^2)$]  and does not involve Delta excitations; see fig.~\ref{fig:2Bdiagrams}. It is the
two-body analogue of the $\pi N$ loop graphs (c) in fig.~\ref{fig:1Bdiagrams},
first computed for $t_{12}=0$ in refs.~\cite{Beane:1999uq, Beane:2004ra} and extended to $t_{12}=1$  in refs.~\cite{Choudhury:2007bh,
  Shukla:2008zc, ShuklaPhD} where full
expressions can be found. These contributions are nonzero only for
$\mathrm{np}$ pairs, \ie~they all contain the same $2\N$ isospin factor of eq.~\eqref{eq:2Bkernelexample}:
\begin{equation}
    \langle t_{12}m^{t\prime}_{12}|(\tau^{(1)} \cdot \tau^{(2)} - \tau^{(1)}_{z} \tau^{(2)}_{z})|t_{12}m^{t}_{12}\rangle=2(-1)^{t_{12}+1}\;
    \delta_{m^{t\prime}_{12}m^t_{12}}\;\delta_{m^t_{12}0}\;\;.
\end{equation} 
Therefore, they parametrise the leading term of both photons hitting the
charged-meson-exchange. In \fourHe, as in \threeHe, both isospin $t_{12}=0$ and $1$
pairs are present. Corrections to these currents enter at one order higher,
\NXLO{4} [$\calO(e^2\delta^4)$], than we consider here.

\subsection{Generation of Transition Densities from Potentials}
\label{sec:potentials}

We used a class of four chiral $2$N and $3$N interactions to generate the one-
and two-body densities for \fourHe: the \ChiEFT Semi-local Momentum-Space (SMS)
$2\N$ potentials in the version ``\NXLO{4}+'' (\ie~considered to be complete at \NXLO{4} with
some \NXLO{5} interactions)~\cite{Reinert:2017usi} and the corresponding
chiral $3\N$ interaction ``\NXLO{2}'' discussed in ref.~\cite{Maris:2020qne}. We employed $2 \N$ momentum-space cutoffs $\Lambda=550\;\MeV$ (hardest),
$500\;\MeV$, $450\;\MeV$ and $400\;\MeV$ (softest) (the complete set of parameters of the 3N interactions can be found in Table~1 of Ref.~\cite{Le:2023bfj}). 
These nuclear interactions all capture the correct long-distance physics of one- and two-pion exchange and
reproduce both the $2\N$ scattering data and the experimental values of the
triton and \threeHe binding energies well. Increasing momentum cutoffs
indicate increasing ``hardness'' of the short-distance interaction. Their \fourHe binding
energies are within $+0.5\;\MeV$ ($\Lambda=550\;\MeV$) to $-0.1\;\MeV$
($\Lambda=400\;\MeV$) of the experimental value. This variation by less than $2\%$
is not a concern, since we compute for photon energies that are high enough that this slight variation in the \fourHe binding energy has no impact on the elastic Compton cross section.

These are of course but four out of a number of modern,
sophisticated potentials. Our choice is dictated by the fact that they are
semi-local (and hence of the form of AV18) and are already coded for the densities formalism. By varying the cutoff within a single family of \ChiEFT potentials, we avoid questions about how the cutoffs of different realisations of the \ChiEFT regulator functions are related. The range of cutoffs chosen, while not large enough to establish renormalisability of the theory, is large enough to indicate a lower bound of the sensitivity of cross sections to the short-distance physics of this process. 


These \ChiEFT wave functions use Weinberg's ``hybrid
approach''~\cite{Weinberg:1990rz}, in which potentials are derived to an assumed
accuracy and then iterated to produce amplitudes or, in our case, one- and
two-body densities. All claim a higher accuracy than that of our Compton
kernels. We refrain here from entering the ongoing debate about correct implementations
of the chiral power counting or the range of cutoff variations, etc.; see
refs.~\cite{Phillips:2013fia, vanKolck:2020llt} for concise summaries, ref.~\cite{Griesshammer:2021zzz} for a polemic, and ref.~\cite{Tews:2022yfb} for a variety of community voices.

Similarly, even though the Compton Ward identities are violated because the
one-pion-exchange $2\N$ potentials are regulated, any inconsistencies between
currents and nuclear densities is compensated by operators which enter at
higher orders in \ChiEFT than the last order we fully retain [\NXLO{3},
$\calO(e^2\delta^3)$]. In addition, the potentials do not include explicit
Delta contributions while the kernel does.  However, it is easy to see that,
for real Compton scattering around $120\;\MeV$, a Delta excited directly by
the incoming photon is more important than one that occurs virtually between
exchanges of virtual pions, especially for an isoscalar target like
\fourHe. For our purposes, such Delta excitations in the $2\N$ potential are
already suppressed by several orders in the chiral counting and well
approximated by the $\pi N$ seagull Low Energy Coefficients that enter the \NXLO{3} interaction.

In sect.~\ref{sec:dependence-pots}, we therefore take the differences between results with the $4$ 
different sets of densities as providing a lower bound indicative of the
present residual theoretical uncertainties. These do not affect the
conclusions of our sensitivity studies, but better extractions of
polarisabilities from \fourHe data will undoubtedly need a reduced potential
spread. 
The results generated with the potential \chiSMSfourfive turn out to be approximately the mean of those generated from the different potentials considered, so we use that for central values, and assess variations with respect to it. 

The \fourHe one- and two-body densities, together with the \threeHe
densities, are publicly available
using the python package provided at~\url{ https://pypi.org/project/nucdens/}. They are defined in momentum space, for centre-of-mass energies and
momentum transfers corresponding to Compton scattering photon energies between
$\omegacm=50$ and $120\;\MeV$ in steps of $10\;\MeV$ and scattering angles
$\thetacm\in[0;180^\circ]$ in steps of $15^\circ$ (momentum-transfers
$\sqrt{\qv^2}\in[0;240]\;\MeV$), plus at selected higher and lower energies
for control. Also available are densities using the harder AV18 $2\N$ model
interaction~\cite{Wiringa:1994wb}, supplemented by the Urbana-IX $3\N$
interaction (3NI)~\cite{Pudliner:1995wk, Pudliner:1997ck}, and densities using the chiral Idaho interaction for
the $2\N$ system in the version ``\NXLO{3}'' at cutoff
$500\;\MeV$~\cite{Entem:2003ft} with the \ChiEFT $3\N$ interaction in the
version ``$\calO(Q^3)$'' using variant ``b'' of
ref.~\cite{Nogga:2005hp} as the ``softest'' choice. 

\section{Results}
\label{sec:results}

\subsection{Overview and Main Result}
\label{sec:over}

\begin{figure}[!ht]
  \begin{center}
  \includegraphics[width=\linewidth]
  {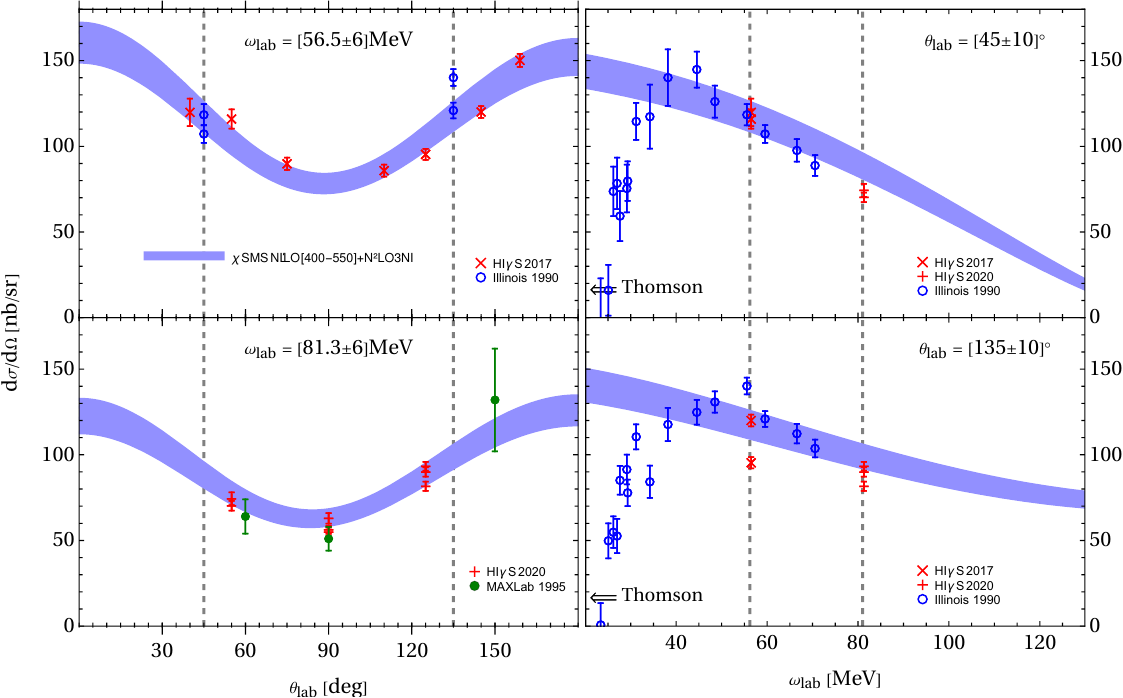}
  
  \caption{(Colour on-line) The band of the differential cross sections at
    \NXLO{3} [$\calO(e^2\delta^3)$]  mapped out by results generated  using the nucleon densities from four $\chi$SMS\NXLO{4}+N$^2$LO3NI potentials with cutoffs
    $\{400;450;500;550\}\;\MeV$, at $\omegalab=56.5\;\MeV$
    (top left) and $81.3\;\MeV$ (bottom left), and angles $\thetalab=45^\circ$
    (top right) and $135^\circ$ (bottom right). All the data from
    Illinois~\cite{WellsPhD}, MAXlab~\cite{Fuhrberg:1995zz} and two \HIGS
    runs~\cite{Sikora:2017rfk, Li:2019irp} are included, via the
    ranges indicated in each plot. (The small effect of the difference
    between nominal and actual energies or angles is not accounted for here; see sect.~\ref{sec:data}
    for details.) Vertical dashed lines in the $\omegalab$
    plots (left) correspond to the angles in the $\thetalab$ plots (right), and
    \emph{vice versa}. On the right, arrows at $20\;\MeV$ indicate the
    Thomson limit ($\omega=0$) at given $\theta$. The disagreement between theory and experiment for $\omegalab\lesssim50\;\MeV$ is expected;  see text for
  details.}
\label{fig:data}

\end{center}
\end{figure}

Figure~\ref{fig:data} summarises our key result, on which we elaborate in the subsequent sub-sections. \ChiEFT agrees well with the available data for $\omegacm\gtrsim50\;\MeV$, as we explain in detail in sect.~\ref{sec:data}.
In sect.~\ref{sec:uncertainties}, we demonstrate that the \ChiEFT
treatment of Compton
scattering on \fourHe at \NXLO{3} [$\calO(e^2 \delta^3)$] has $\lesssim \pm10\%$  uncertainties 
from the dependence on the $2\N$ and $3\N$ interactions employed and from an assessment of order-by-order convergence of the EFT. 

\begin{figure}[!htbp]
\begin{center} \includegraphics[width=0.5\linewidth]
  {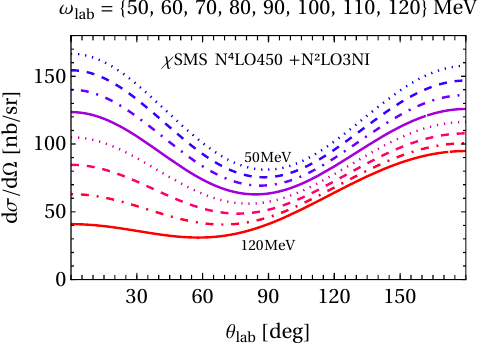}
  \caption{(Colour on-line) Evolution of the differential \fourHe Compton
    cross section at \NXLO{3} [$\calO(e^2\delta^3)$], in the lab frame from $50$ (top) to
    $120\;\MeV$ (bottom) in steps of $10\;\MeV$, for densities generated using
    the ``mean'' potential \chiSMSfourfive.}
\label{fig:energies}
\end{center}
\end{figure}

Figure~\ref{fig:energies} shows that the cross section
drops steadily with energy. Concurrently, it starts out almost fore-back
symmetric but becomes quite lopsided as scattering under backward angles grows
relative to forward angles. At $120\;\MeV$, back-angle cross sections are
about twice as big as forward-angle ones and have reduced to a bit more than
half of their $\omegalab=50\;\MeV$ values. 
This is consistent with findings for the
deuteron and \threeHe~\cite{Hildebrandt:2004hh, Hildebrandt:2005ix, Hildebrandt:2005iw,
  Griesshammer:2013vga, Margaryan:2018opu}, although there the fore-aft asymmetry is with only about $1:1.2$ not nearly as dramatic.

We analyse the sensitivity to the scalar-isoscalar dipole polarisabilities of the nucleon in sect.~\ref{sec:polarisabilities}.
We will concentrate on three energies
$\{60;90;120\}\;\MeV$ to cover the possible range of future experiments. For
these, the subsequent figures~\ref{fig:orders} to~\ref{fig:polarisabilities}
at $\{45^\circ;120^\circ;150^\circ\}$ display the energy dependence of our results. We consider these to provide a
good compromise between experimental feasibility and valuable information on
the polarisabilities. In sect.~\ref{sec:beamasymmetry}, the beam asymmetry's convergence and
sensitivity to $\alphaes$ and $\betams$ is addressed.

Throughout secs.~\ref{sec:uncertainties} to \ref{sec:beamasymmetry}  we assess the difference between two quantities (\eg~the cross
section using two different potentials) by defining:
\begin{equation}
  \label{eq:reldev}
  \mbox{``relative deviation of $A$ from $M$'' } :=\frac{A}{M}-1\;\;. 
\end{equation}

\subsection{Comparison with Data}
\label{sec:data}

A total of $50$ measurements are available. Those with the highest precision
come from \HIGS using a quasi-monchromatic photon beam: $7$ points at
$56.5\;\MeV$ published in 2017~\cite{Sikora:2017rfk}\footnote{While all other experiments report the average beam energy, this one only quotes $61\;\MeV$ as the energy of the maximum beam intensity. Due to an asymmetric beam energy profile, this translates into a weighted mean energy of $56.5\;\MeV$ which we use~\cite{Sikora}. We thank M.~H.~Sikora for providing these details.}, and another $8$ at $81.3\;\MeV$
published in 2020~\cite{Li:2019irp}. Older data are available from two tagged-photon
bremsstrahlung facilities: $3$ points from MAXlab at $87\;\MeV$ in
1995~\cite{Fuhrberg:1995zz}; and the largest dataset, $32$ points, from the
University of Illinois Tagged Photon Facility between $23.4\;\MeV$ and
$70.55\;\MeV$ at $45^\circ$ and $135^\circ$~\cite{WellsPhD}. This pioneering
set from 1990 covers the widest energy range, but also carries the largest
uncertainties. We present the data with error bars which add statistical and
point-to-point systematic uncertainties in quadrature but which do not account
for systematic correlated/overall uncertainties. These are reported as $2.2\%$
for both \HIGS sets, $15\%$ for the MAXlab set, but not reported for the Illinois
set.

To compare these data with our result in fig.~\ref{fig:data}, we chose $4$
fixed-energy and fixed-angle slices in which each datum is represented at least once: two fixed-energy
plots at the energies of the \HIGS data, $56.5\;\MeV$ and $81\;\MeV$, but
containing also data within $\pm6\;\MeV$ of the central values; and two
fixed-angle plots at the angles of the Illinois data---$45^\circ$ and
$135^\circ$---but including data within $\pm10^\circ$. Cognisant of the present
accuracies of both data and theory, we do not correct cross sections of data
whose energy or scattering angle do not exactly match the central values; the differences are visible but smaller than even the experimental error bars. 
The Illinois data for $56.5\;\MeV$ at back-angles appear somewhat higher than
the \HIGS ones, while all are consistent at forward angles. At $81.3\;\MeV$,
all data appear to be consistent. At fixed angles, the Illinois set appears to
be slightly more sloped as a function of energy than the \HIGS data and the
theory curve. It may, however, be worth recalling that the Illinois data were
obtained at a first-generation tagged-photon facility. 

Agreement of theory and data is good within the
experimental and theoretical uncertainties \emph{in the range where the
  assumptions of the present theoretical description hold}: namely for those
$\omega\sim\mpi$ for which the intermediate four-nucleon system predominantly
propagates incoherently, with only minor effects from restoring the Thomson
limit, as discussed in sect.~\ref{sec:kernels}. At $\omega\lesssim50\;\MeV$,
the only data are from Illinois~\cite{WellsPhD}. They show a rapid drop
towards the Thomson limit after a ``knee'' around $40$ to $50\;\MeV$. These data
at the lowest energies are not inconsistent with the Thomson limit (denoted by
arrows at $20\;\MeV$ in the energy-dependent plots). Informed by these comparisons
and accounting for the uncertainties of both theory and experiment, 
we conclude that the incoherent-propagation assumption is
justified for $\omega\gtrsim50\;\MeV$, consistent with the
\emph{a-priori} estimate of sect.~\ref{sec:apriori}. 

To put their \HIGS data in context, Sikora \etal~ used a phenomenological model to parametrise the \fourHe elastic Compton cross section~\cite{Sikora:2017rfk}. In it, the density of nucleons is captured as a Fermi function; photons interact with the nucleons through one-body $E1,E1$ and $M1,M1$ operators; and two-body currents are included so that they provide the right overall strength. Sikora \etal's isoscalar polarisability values of $10.9$ and $3.6$ are close to the numbers employed here. Crucially though, the one-body amplitude does not include dispersive effects in $\alpha_{E1}$ and $\beta_{M1}$. However, the energy dependence of these polarisabilities is known to be a crucial effect for photon energies approaching $100\;\MeV$~\cite{Hildebrandt:2003fm, Hildebrandt:2005ix, Hildebrandt:2005iw, Griesshammer:2012we}; this is likely responsible for the disagreement between this model and their data. In addition, it has been demonstrated that higher photon multipoles are relevant at such energies in the $\gamma$-deuteron reaction~\cite{Hildebrandt:2005ix,
  Hildebrandt:2005iw}. All this stems from the fact that Sikora \etal's model does not account for the pion dynamics in either the polarisabilities or the exchange currents, and so lacks a microscopic description of the energy dependence of the former and the size of the latter. In contrast, our \ChiEFT calculation predicts both of these effects via a consistent underlying theory. It also delivers a much more accurate treatment of the \fourHe wave function.

The systematic nature of the \ChiEFT expansion means that we can estimate the uncertainties induced by truncating its $\gamma {}^4$He amplitude at \NXLO{3} [$\calO(e^2\delta^3)]$---something that is required for a
 rigorous comparison between \ChiEFT and these data. We will discuss in the next subsection how that uncertainty estimate produces the bands in fig.~\ref{fig:data}.

\subsection{Theoretical Uncertainties}
\label{sec:uncertainties}

As in previous presentations on Compton scattering on the deuteron~\cite{Griesshammer:2012we, Griesshammer:2015ahu} and
\threeHe~\cite{Margaryan:2018opu}, we assess convergence and theoretical
uncertainties in three ways, each based on the \ChiEFT expansion:
\emph{a-priori}; order-by-order; and residual dependence on short-distance
details.  As described in sect.~\ref{sec:kernels}, the Compton kernel is
incomplete at low energies, $\omega\lesssim50\;\MeV$ because rescattering
contributions become important and finally restore the correct Thomson limit
as $\omega\to0$. Thus, the estimates we are about to discuss do not apply
there.

We reiterate that our results are exploratory and do not yet suffice for
high-accuracy extractions of polarisabilities from data. For that, the Thomson
limit should be restored. This should also reduce the dependence on the $2\N$
and $3\N$ interaction. Therefore, we do not explore a rigorous interpretation
of theoretical uncertainties based on Bayesian statistics as developed
in~\cite{Furnstahl:2015rha,Griesshammer:2015ahu,Melendez:2019izc,Melendez:2020ikd}.

\subsubsection{\emph{A-Priori} Estimate}
\label{sec:apriori}

For the \emph{a-priori} estimate, the Compton amplitudes are complete up to and
including \NXLO{3} [$\calO(e^2\delta^3)$] and therefore carry an uncertainty
of roughly $\delta^4\approx(0.4)^4\approx\pm3\%$ of the LO
result. This translates into about $\pm6\%$
for cross sections and beam asymmetries since these are proportional to the square of
amplitudes. This appears to be a slight
under-estimate when placed alongside the following two \emph{post-facto} criteria, but
the contribution at any EFT order is always multiplied by a number of order one,
and there is also some uncertainty in the breakdown scale. 

\subsubsection{Convergence of the \texorpdfstring{\ChiEFT}{ChiEFT} Expansion}
\label{sec:convergence}

\begin{figure}[!th]
\begin{center} \includegraphics[width=\linewidth]
  {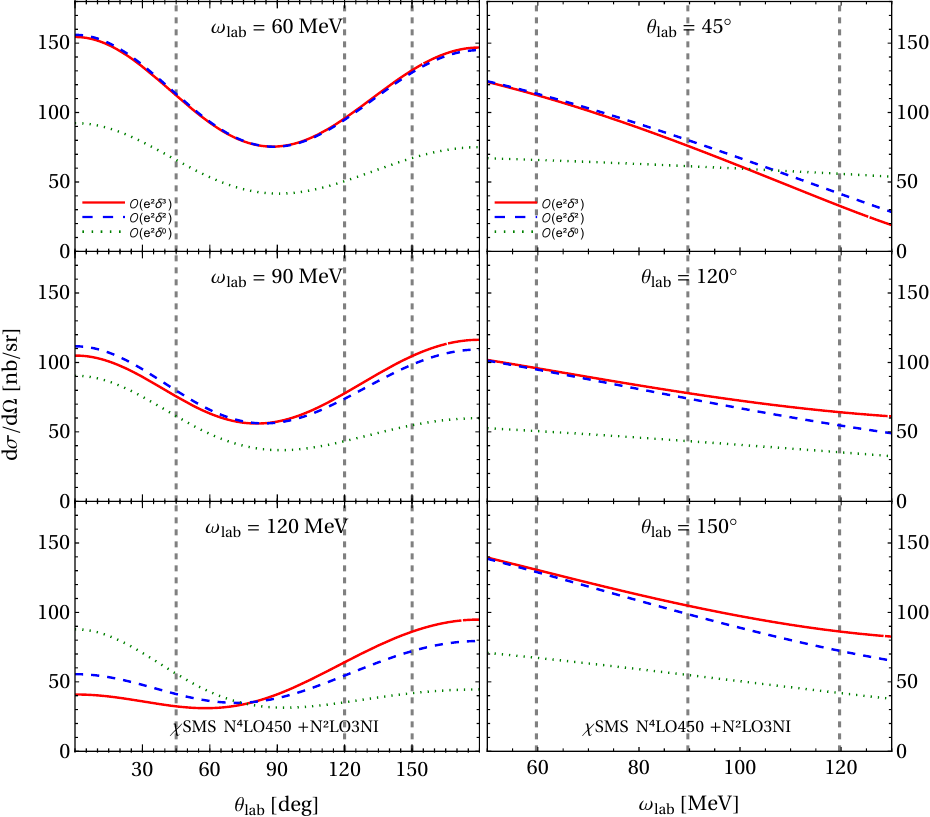}
  \caption{(Colour on-line) The cross section at \NXLO{3} [$\calO(e^2\delta^3)$] (with
    Delta; red solid), \NXLO{2} [$\calO(e^2\delta^2)$] (no Delta; blue dashed), and
    LO [$\calO(e^2\delta^0)$] (single-nucleon Thomson term only; green dotted), using the
    ``mean'' potential \chiSMSfourfive; see text for details.  Vertical dashed lines in
    the $\omegalab$ plots (left) correspond to the angles in the $\thetalab$
    plots (right), and \emph{vice versa}.}
\label{fig:orders}
\end{center}
\end{figure}

The second assessment uses order-by-order convergence. So that
differences between the orders are not contaminated by spurious dependencies, we use the same values for the static nucleon
polarisabilities at both \NXLO{2} [$\calO(e^2\delta^2)$] and \NXLO{3} [$\calO(e^2\delta^3)$] (no
polarisabilities enter at LO [$\calO(e^2\delta^0)$]). We notice that the sensitivity of
observables to varying polarisabilities is for all practical purposes
unaffected by the exact choice of their central values.

As fig.~\ref{fig:orders} shows, the correction from $\calO(e^2\delta^2)$
[\NXLO{2}, \ie~without the Delta] to $\calO(e^2\delta^3)$ [\NXLO{3},
\ie~including the Delta] is indeed smaller than that from $\calO(e^2\delta^0)$
[LO] to $\calO(e^2\delta^2)$ [\NXLO{2}]; remember that there is no correction
at $\calO(e^2\delta^1)$ [NLO]. That the LO-to-\NXLO{2} correction is generally
quite large in Compton scattering has already been observed for the nucleon,
deuteron, and \threeHe~\cite{Margaryan:2018opu}. After all, LO is just the
Thomson term for the proton, and the contributions from
charged-pion-exchange currents are significant in the
deuteron and \threeHe~\cite{Hildebrandt:2004hh, Hildebrandt:2005ix,
  Hildebrandt:2005iw, Griesshammer:2013vga, Margaryan:2018opu}. While a correction
of about $50\%$ at $60\;\MeV$ is similar to that in
\threeHe~\cite{Margaryan:2018opu}, the correction of about $50\%$ at
$120\;\MeV$ is smaller than the $ \lesssim70\%$ correction in \threeHe,
especially at forward angles. This can be explained by a combination of
effects. First, \fourHe is more tightly bound and hence smaller. Second, and perhaps more important, in the isospinor target \threeHe, both the rather small isoscalar
magnetic moment $\mu^{(\mathrm{s})}\approx0.43$ and the much larger isovector
piece $\mu^{(\mathrm{v})}\approx 4.7$ contributes. Since \fourHe is isoscalar, so
the latter is absent at the order we consider.
Overall, the size of the LO-to-\NXLO{2} correction is a poor predictor for the
typical size of higher-order terms. We will not use it to estimate
uncertainties based on convergence.

On the other hand, the correction from $\calO(e^2\delta^2)$ [\NXLO{2}] to
$\calO(e^2\delta^3)$ [\NXLO{3}] becomes visible only at energies
above about $80\;\MeV$. This is not entirely surprising since the additional
pieces contain the $\Delta(1232)$ whose effect beyond that on the static polarisabilities is constrained be small at low
energies. Further charged-meson-exchange contributions at \NXLO{4} will shed
more light on whether such a small shift is 
accidental. Towards
$120\;\MeV$, the $\Delta(1232)$'s effect changes the angle dependence
significantly, reducing forward scattering by about $35\%$ while increasing
back-scattering by about $20\%$. That is consistent with the findings in the
deuteron and \threeHe~\cite{Hildebrandt:2005ix,
  Hildebrandt:2005iw, Griesshammer:2013vga, Margaryan:2018opu}. It might at first be surprising
that the $\Delta(1232)$ has such a large effect well below its resonance
threshold. However, ref.~\cite{Margaryan:2018opu} already pointed out that while
the static magnetic polarisability $\betams$ is by construction unchanged between orders,
the $\Delta(1232)$ adds a sizeable dispersive correction which can be as large
as the static value of $\betams$ itself; see the discussion of ``dynamical
polarisabilities'' in refs.~\cite{Griesshammer:2012we, Griesshammer:2017txw}.

Taking all his into consideration, we take about $\delta\approx0.4$ of the 
difference between \NXLO{2} and \NXLO{3} itself as total width of the
uncertainty band from higher-order corrections. That amounts to just a
per-cent at $\omega\approx60\;\MeV$, which is clearly an under-estimate also in view of the \emph{a-priori} estimate above. It increases for $120\;\MeV$ to about $\pm8\%$ forward and about $\pm4\%$
backward, or some $\pm6\%$ overall.

\subsubsection{Residual Dependence on the \texorpdfstring{$2\N$}{2N} and \texorpdfstring{$3\N$}{3N} Interactions}
\label{sec:dependence-pots}

\begin{figure}[!b]
\begin{center} \includegraphics[width=\linewidth]
  {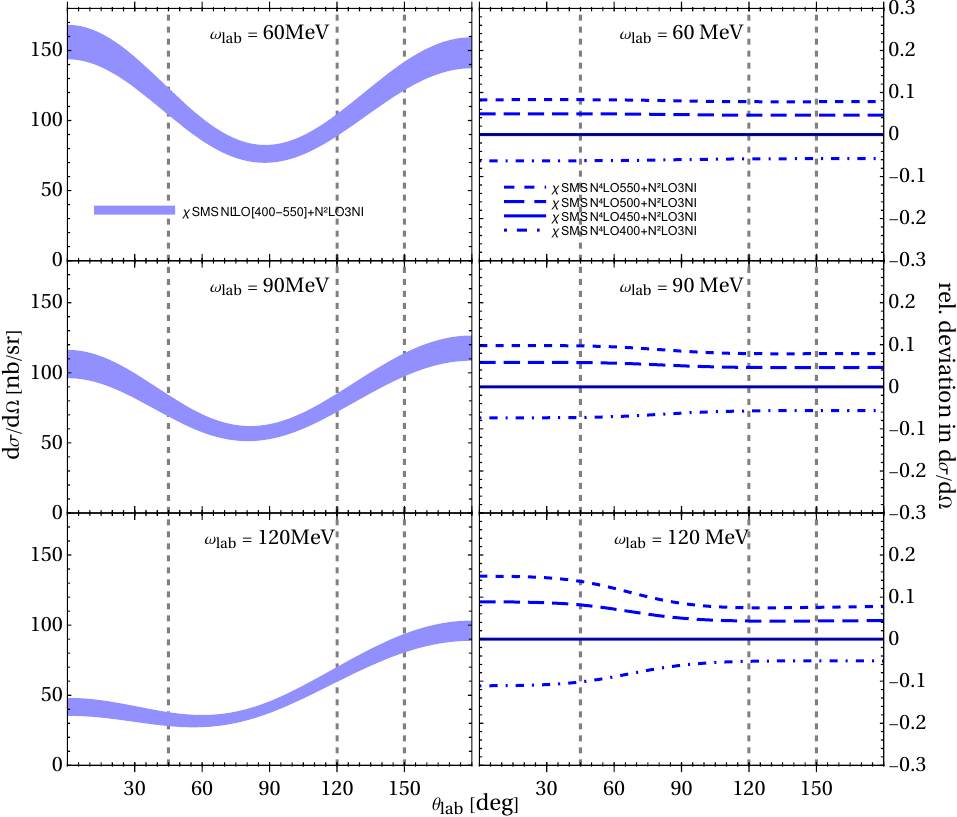}
  \caption{(Colour on-line) Energy- and angle-dependence of 
    the four $\chi$SMS\NXLO{4}+N$^2$LO3NI potentials with cutoffs
    $\{400;450;500;550\}\;\MeV$ at \NXLO{3} [$\calO(e^2\delta^3)$].
    Left: spread of the potentials, using a range identical to that in the relative-sensitivity plot on the left of fig.~\ref{fig:polarisabilities}, to aid comparison. 
    Right: relative difference of each potential to a ``mean'' potential
    \chiSMSfourfive at \NXLO{3} [$\calO(e^2\delta^3)$]; see text for details. In both panels,
    vertical dashed lines denote the angles $\thetalab=\{45^\circ;120^\circ;150^\circ\}$ of
    fig.~\ref{fig:orders}.}
\label{fig:potentials}
\end{center}
\end{figure}

As discussed in sect.~\ref{sec:potentials}, the $\chi$SMS\NXLO{4}+N$^2$LO3NI
$2\N$ and $3\N$ potentials used to produce the one- and two-body densities all
capture the correct long-range physics of one- and two-pion exchange and
reproduce the two-nucleon scattering data and the \threeHe and \fourHe binding
energies equally well---indeed, at a level superior to the accuracy aimed for in this
article. Therefore, differences only stem from the different cutoffs
$\{400;450;500;550\}\;\MeV$, \ie~from short-range dynamics not explicitly 
included in \ChiEFT.  Those differences therefore
provide an estimate of the effect of higher-order terms in the \ChiEFT
series. This estimate provides a lower bound of the
theoretical uncertainties.

Figure~\ref{fig:potentials} shows the bands mapped out by the four cutoffs as
well as the relative deviation from the ``mean'' potential \chiSMSfourfive,
indicating for comparison also the angles
$\thetalab=\{45^\circ;120^\circ;150^\circ\}$ of fig.~\ref{fig:orders}. Since a
larger cutoff translates to a harder
interaction, one concludes that
softer interactions correlate with smaller cross sections.
The relative width is practically
angle-independent at $\omegalab=60\;\MeV$. With growing energy, it remains
largely constant for back-angles at $\pm7\%$ but increases for forward angles
to about $\pm12\%$ at $120\;\MeV$, with the increase roughly proportional to
$\omegacm^2$. Meanwhile, the absolute width of the band decreases somewhat for higher energies, but this is due to the drop in the cross section with increasing $\omega$ .

As an aside, results with the relatively ``hard'' AV18 $2\N$ model
interaction~\cite{Wiringa:1994wb}, supplemented by the Urbana-IX $3\N$
interaction~\cite{Pudliner:1995wk, Pudliner:1997ck}, add angle-independently about $2\%$ to the
upper limit of the band at low energies, and about $3\%$ at $120\;\MeV$. On
the other hand, the much ``softer'' chiral Idaho $2\N$ interaction
``\NXLO{3}'' at cutoff $500\;\MeV$~\cite{Entem:2003ft} with the $3\N$
interaction ``$\calO(Q^3)$'' of variant ``b'' of
ref.~\cite{Nogga:2005hp} produces cross sections which are substantially lower
than the lower limit of the bands, namely adding about $-4\%$ (forward) and
$-6\%$ (backward) at low energies, culminating in about $-15\%$ across all
angles at high energies. These two potentials map out extremes of similar relative
sizes in \threeHe~\cite{Margaryan:2018opu}. 

  
\subsubsection{Numerical Uncertainties}
\label{sec:numerics}

Numerically, all results are fully converged in the radial and angular
integrations to relative deviations of better than $\pm0.4\%$ at the highest
energies and momentum-transfers $\qv^2$ (\ie~large $\thetalab$ in~eq.~\eqref{eq:momtransfer}), with
significantly smaller uncertainties at low energy and/or low momentum transfer. In
particular, we keep all partial waves $j_{12}\le2$ in the two-body matrix
elements. Test computations using $j_{12}\le3$ indicate that contributions
from higher partial waves change the two-body matrix elements by at most
$0.2\%$ across all energies and angles. These uncertainties are therefore
negligible compared to the truncation uncertainties. 

\subsubsection{Overall Estimate of Theoretical Uncertainties}
\label{sec:estimate}

We are therefore provided with three error assessments: \emph{a priori}
$\pm6\%$; from order-by-order convergence with a percent at low energies and
$\pm8\%$ (forward) and $\pm4\%$ (backward) at higher energies; and a residual cutoff dependence
of $\pm7\%$ overall that does reach $\pm12\%$ at $120\;\MeV$ and forward angles.  These
are all consistent with each other, and the residual cutoff dependence appears
to provide the biggest uncertainties. We are therefore comfortable -- for the
purpose of this first study of \fourHe Compton scattering -- to conservatively
assign an overall theory uncertainty of $\pm10\%$ at all energies and
angles, plus possibly an additional $\pm2\%$ for the back-angles at high
energies. This is also consistent with our analogous estimate of $\pm10\%$ uncertainties in \threeHe Compton scattering~\cite{Margaryan:2018opu}. We defer a more thorough estimate using a rigorous interpretation of
theoretical uncertainties based on Bayesian statistics and Gaussian process modeling
of the angle and energy dependence~\cite{Melendez:2019izc,Melendez:2020ikd}
to a future publication extracting nucleon polarisabilities from \fourHe
Compton scattering.

\subsection{Sensitivity to the Scalar-Isoscalar Polarisabilities}
\label{sec:polarisabilities}

\begin{figure}[!b]
  \begin{center}

    \includegraphics[width=\linewidth]{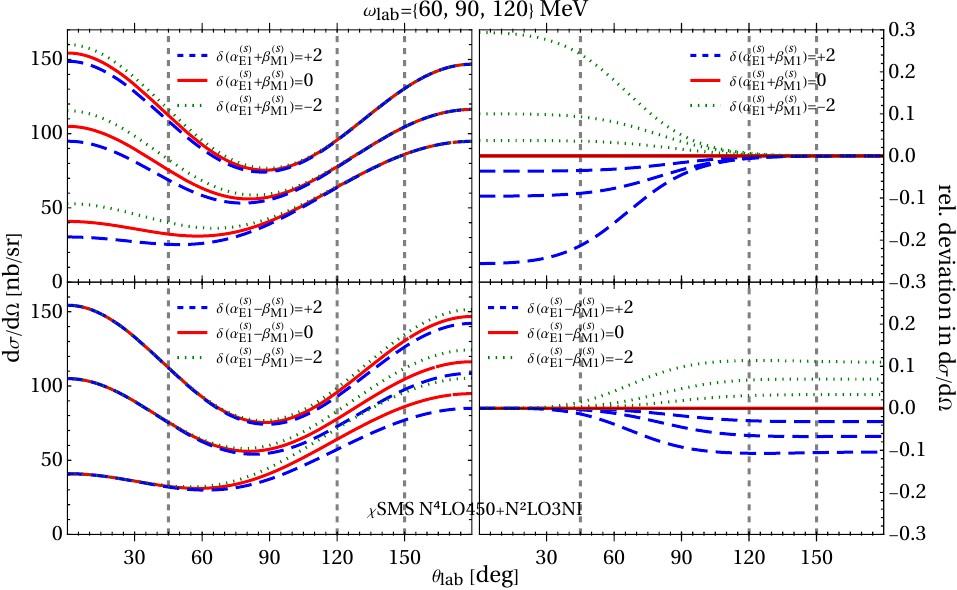}
    
  \caption{(Colour on-line) Sensitivity of the cross section to varying the
    scalar-isoscalar polarisabilities around their central values (solid line)
    of eq.~\eqref{eq:alphabeta} by $+2$ (blue dashed) and $-2$ (green dotted)
    units,
    at \NXLO{3} [$\calO(e^2\delta^3)$] using the ``mean'' potential \chiSMSfourfive; see
    text for details. Left: Impact on the differential cross section. Right:
    Relative deviation 
    from the central value
    at $60$, $90$ and $120\;\MeV$ (innermost to outermost in each panel). 
    At all energies, sensitivities to isovector or spin polarisabilities are
    practically absent. While other potentials change the overall size of the cross section by 
    $\pm 10\%$
    (see fig.~\ref{fig:potentials}), relative
    sensitivities to scalar-isoscalar polarisabilities remain near-identical. Vertical dashed lines
    denote the angles $\thetalab=\{45^\circ;120^\circ;150^\circ\}$ of
    figs.~\ref{fig:orders} and~\ref{fig:potentials}.}
\label{fig:polarisabilities}
\end{center}
\end{figure}

As in previous studies~\cite{Hildebrandt:2005ix,
  Hildebrandt:2005iw, Griesshammer:2013vga, Margaryan:2018opu}, we choose a variation of the
scalar-isoscalar polarisabilities by $\pm2$ canonical units in
fig.~\ref{fig:polarisabilities}. This is roughly at the level of the combined
statistical, theoretical and Baldin Sum Rule induced uncertainties of the
scalar polarisabilities. As \fourHe is a scalar-isoscalar target, this corresponds 
to varying the corresponding neutron polarisability by $\pm4$. While this is twice the variation 
we had considered for \threeHe~\cite{Margaryan:2018opu}, it aligns with the different linear 
combinations of proton and neutron polarisabilities accessible in these isotopes. 
Other variations are very well-determined by linear
extrapolation from our results, since quadratic contributions of the
polarisability variations $\delta(\alphaes,\betams)$ are suppressed in the
squared amplitudes. 

In addition, we checked that spin and isovector-scalar
polarisabilities leave no signal in \fourHe Compton observables, as expected
for a perfect scalar-isoscalar target. It is worth reiterating that an extraction from deuteron and \threeHe data is
potentially contaminated by uncertainties in the spin-isoscalar
polarisabilities, most notably in $\gamma_{E1E1}^{(\mathrm{s})}$, while
extractions on \fourHe will not suffer from such issues. 

Since the cross section is only sensitive to $\alphaes+\betams$ at forward
angles, there one can check there consistency of data with the Baldin Sum Rule value
of $14.5\pm0.4$~\cite{OlmosdeLeon:2001zn, Levchuk:1999zy}. More interesting is
the combination $\alphaes-\betams$ which is at present much less accurately
determined as $7.8\pm1.2(\mathrm{stat})\pm0.8(\mathrm{th})$ from the world
deuteron data~\cite{Griesshammer:2012we, Myers:2014ace, Myers:2015aba};
\cf~eq.~\eqref{eq:alphabeta}. Note that  the combined error of $\pm1.5$ is
of the order of the $\pm 2$ variations considered here.

Figure~\ref{fig:polarisabilities} illustrates again the familiar
trade-off: With increasing energy, the cross section (and hence the event
rate) decreases but the sensitivity to the polarisabilities increases. The
difference induced by varying the scalar-isoscalar polarisabilities at
$\omegalab=60\;\MeV$ by $\pm2$ canonical units is $\pm3\%$ and so just as much as 
the thickness of the line, amounting to at best $\pm4\;\mathrm{nb/sr}$ 
at the most extreme angles. At $120\;\MeV$, however, a variation of the
crucial parameter $\alphaes-\betams$ by $\pm2$ changed the cross section by a near-constant $\pm11\%$ between $\thetalab\approx100^\circ$ and $180^\circ$. That
translates, for example, to about $\pm9\;\mathrm{nb/sr}$ at a realistically
accessible back-angle of $150^\circ$. Since the forward cross section is about
half of the back-angle one, the relative sensitivity to $\alphaes+\betams$ is
about twice that of $\alphaes-\betams$; the sensitivity in the rates is
comparable.

It must be emphasised that both the absolute and relative \emph{sensitivities}
to polarisability variations are typically very little affected by the
theoretical uncertainties discussed in the preceding subsection. The effect of
the discrepancies between different potentials is mitigated by the fact that a
large part of that relative deviation is angle-independent, whereas the
sensitivities to the scalar-isoscalar polarisabilities $\alphaes\pm\betams$
have a rather strong angular dependence. We therefore judge that our
sensitivity investigations are sufficiently reliable to be useful for current
planning of experiments---as we previously argued for
\threeHe~\cite{Margaryan:2018opu}. We reiterate that our goal here is an
exploratory study of magnitudes and sensitivities to the nucleon
polarisabilities. A polarisability extraction will of course need to address
residual theoretical uncertainties with more diligence, as was already done
for the proton and deuteron in refs.~\cite{Griesshammer:2012we,
  McGovern:2012ew, Myers:2014ace, Myers:2015aba, Griesshammer:2015ahu,
  Griesshammer:2017txw, Melendez:2020ikd}.

We therefore propose that data at small angles and energies should be paired
with data at larger angles and energies. The former provide important checks
on data consistency with the Thomson limit and with the Baldin Sum Rule, while
the latter provides the best chance for high-accuracy extractions of the
scalar-isoscalar polarisability combination $\alphaes-\betams$. Given the
rather straightforward sensitivity to polarisabilities in the validity range of our theory,
$\omegalab\in[50;120]\;\MeV$, we are confident a more sophisticated study using Bayesian
experimental design~\cite{Melendez:2020ikd} will arrive at the same result.

\subsection{The Beam Asymmetry}
\label{sec:beamasymmetry}

Compton beam asymmetries can be measured at some facilities, and data
are available for the proton~\cite{Blanpied:1996, 
  Blanpied:2001, Martel-PhD:2013, Sokhoyan:2016yrc, Collicott:2015,
  Martel:2017pln, HIGS2019, A2:2019bqm,A2CollaborationatMAMI:2021vfy}. They are dominated by the
single-nucleon Thomson term, which for a charged particle at $\omega=0$
predicts a shape $(\cos^2\theta-1)/(\cos^2\theta+1)$ with extremum $-1$ at
$\thetalab=90^\circ$, as is well-known from Classical
Electrodynamics~\cite{Jackson}.
In our findings for \fourHe, we do not address uncertainties
and sensitivities at the extreme angles $\thetalab\le10^\circ$ and
$\thetalab\ge160^\circ$ since the asymmetry is quite small ($\le0.1$) there.

\begin{figure}[!b]
\begin{center}
     \includegraphics[width=0.5\linewidth]{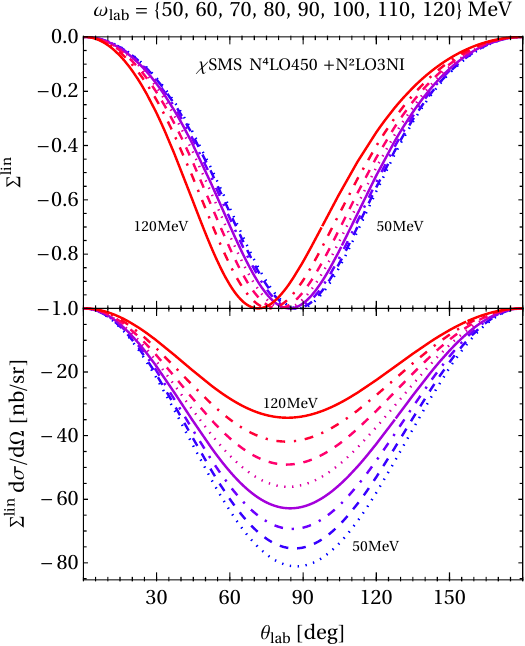}
     \caption{(Colour on-line) Evolution of the \fourHe Compton beam asymmetry
       (top) and its associated cross section difference (bottom)
       at \NXLO{3} [$\calO(e^2\delta^3)$], in the lab frame from $50$ (rightmost minimum)
       to $120\;\MeV$ (leftmost minimum) in steps of $10\;\MeV$, for densities
       generated using the ``mean'' potential \chiSMSfourfive.}
\label{fig:beamasym-energies}
\end{center}
\end{figure}

\begin{figure}[!b]
\begin{center}
     \includegraphics[width=\linewidth]{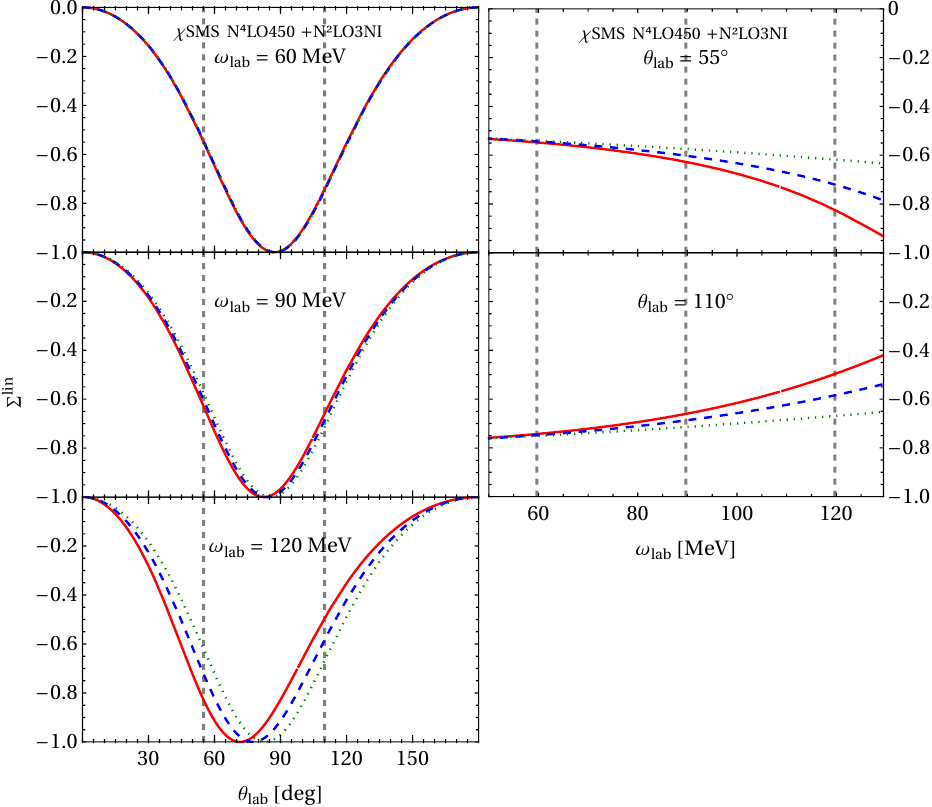}
\caption{(Colour on-line)  The beam asymmetry at \NXLO{3} [$\calO(e^2\delta^3)$] (with
    Delta; red solid), \NXLO{2} [$\calO(e^2\delta^2)$] (no Delta; blue dashed), and
    LO [$\calO(e^2\delta^0)$] (single-nucleon Thomson term only; green dotted), using the
    ``mean'' potential \chiSMSfourfive; see text for details.  Vertical lines in
    the $\omegalab$ plots (left) correspond to the angles in the $\thetalab$
    plots (right), and \emph{vice versa}.}
\label{fig:beamasymmetry}
\end{center}
\end{figure}

\begin{figure}[!b]
\begin{center} \includegraphics[width=0.5\linewidth]
  {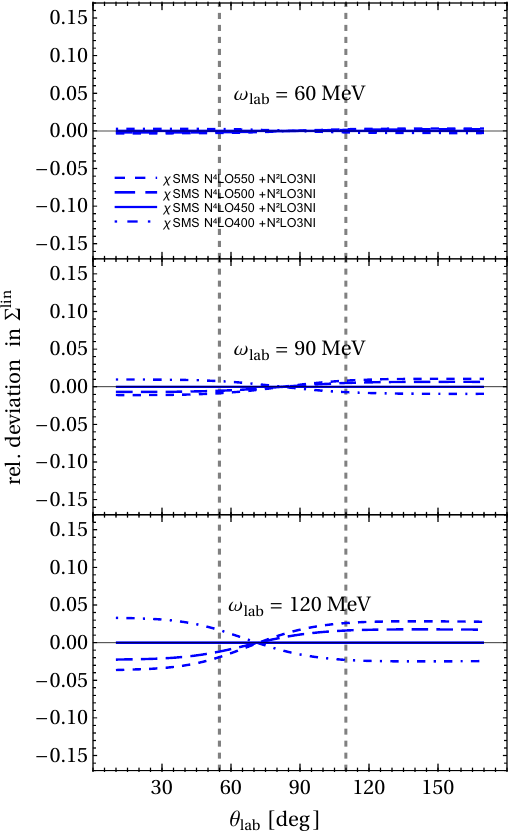}
  \caption{(Colour on-line) Energy- and angle-dependence of relative
    differences of the beam asymmetry for each potential, compared to the
    ``mean'' potential \chiSMSfourfive at \NXLO{3} [$\calO(e^2\delta^3)$]; see text for
    details. The range is identical to that in the relative-sensitivity plot, the right-hand panel of fig.~\ref{fig:beamasym-pols}, to aid comparison. Vertical dashed lines denote the
    angles $\thetalab=\{55^\circ;110^\circ\}$ of
    fig.~\ref{fig:beamasymmetry}. Angles
    $\thetalab\le10^\circ$ and $\thetalab\ge160^\circ$ are not shown since
    they suffer from small rates and large uncertainties.}
\label{fig:beamasym-pots}
\end{center}
\end{figure}

The top panel in fig.~\ref{fig:beamasym-energies} shows the beam asymmetry
itself; the bottom one half of the rate-difference\footnote{In ref.~\cite{Margaryan:2018opu}, $\dd\sigma^\parallel-\dd\sigma^\perp$ is denoted by
  $\Delta_3$.}, $\half(\dd\sigma^\parallel-\dd\sigma^\perp)$. While the rate-difference reduces by two-thirds from $50\;\MeV$
to $120\;\MeV$, the magnitude and shape of the (Thomson-term) asymmetry at zero energy is
largely retained even between $\omegalab=50\;\MeV$ and $120\;\MeV$, with the
extremum merely drifting towards smaller $\thetalab$. As an analysis of the
order-by-order convergence in fig.~\ref{fig:beamasymmetry} shows, this is in
part a simple recoil effect of the Thomson term [LO, $\calO(e^2\delta^0)$],
but the nucleons' magnetic moments and the nucleus' meson-exchange structure change the
position of the minimum by an equal amount of about $5\%$ at the highest energies
[\NXLO{2}, $\calO(e^2\delta^2)$]. Including the $\Delta(1232)$ at \NXLO{3}
[$\calO(e^2\delta^3)$] shifts it yet again by about the same amount, mostly
via the energy-dependent effects in $\betams$. At
$\thetalab=\{55^\circ;110^\circ\}$, the change from \NXLO{2} to \NXLO{3}
suggests an uncertainty of about $\pm2\%$.  The dependence on the $\chi$SMS
$2\N$ and $3\N$ interactions in fig.~\ref{fig:beamasym-pots} is quite small, namely less than thrice the thickness of the lines, and 
rising from a mere per-cent at $60\;\MeV$ to $\pm4\%$ at $120\;\MeV$. Results
for AV18+UIX are near-identical to \chiSMSfivefive; for Idaho
N$^3$LO500+3NFb, they are indistinguishable from \chiSMSfour except at
$120\;\MeV$ where this much softer interaction adds about $+3\%$ at $30^\circ$ and
$-2\%$ at $\gtrsim120^\circ$. Following our discussion in
sect.~\ref{sec:uncertainties}, it seems appropriate to assume a theoretical
uncertainty which rises from $\pm1\%$ at $50\;\MeV$ to $\pm5\%$ at
$120\;\MeV$.  All this indicates rapid convergence in
\ChiEFT. 

\begin{figure}[!b]
\begin{center}
     \includegraphics[width=\linewidth]{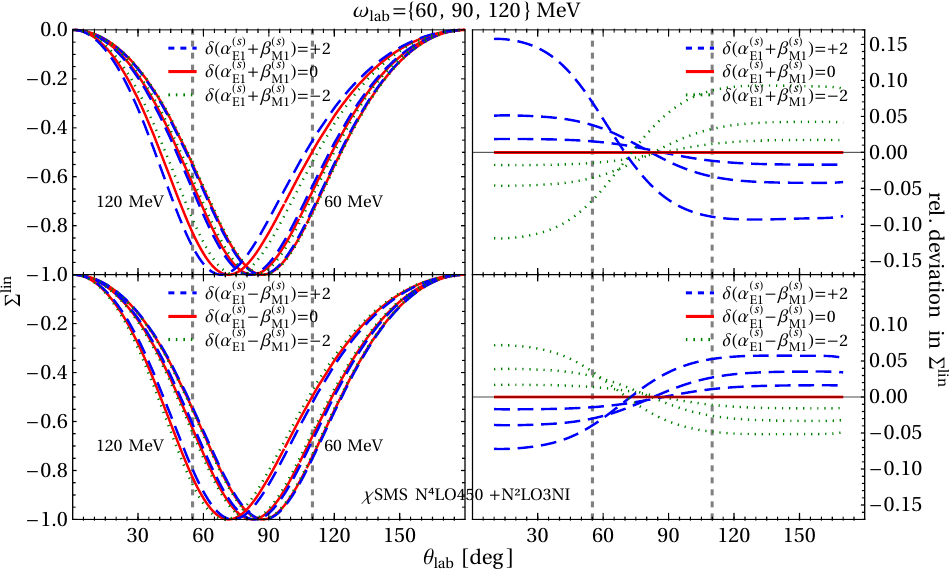}
     \caption{(Colour on-line) Sensitivity of the beam asymmetry to varying the
    scalar-isoscalar polarisabilities around their central values (solid line)
    of eq.~\eqref{eq:alphabeta} by $+2$ (blue dashed) and $-2$ (green dotted)
    units,
    at \NXLO{3} [$\calO(e^2\delta^3)$] using the ``mean'' potential \chiSMSfourfive; see
    text for details. Left: Impact on the beam asymmetry, with the minimum of $\Sigma^\text{lin}$
       wandering to the left as $\omegalab$ increases. Right:
    Relative deviation 
    from the central value
    at $60$, $90$ and $120\;\MeV$ (innermost to outermost in each panel). 
    While other potentials scale the results by up to
    $\pm4\%$ (see fig.~\ref{fig:beamasym-pots}), the relative overall
    sensitivities remain near-identical; see text for details. Vertical dashed lines
    denote the angles $\thetalab=\{55^\circ;110^\circ\}$ of
    fig.~\ref{fig:beamasymmetry}. Relative deviations for
    $\thetalab\le10^\circ$ and $\thetalab\ge160^\circ$ are not shown since
    they suffer from small rates and large uncertainties.}
\label{fig:beamasym-pols}
\end{center}
\end{figure}

Given the persistence of the point-like behaviour, it should come as no
surprise that the relative sensitivity to variations of the polarisabilities
by $\pm2$ units is about half of that for the cross section at the same
energy; see fig.~\ref{fig:beamasym-pols}. Still, since measuring asymmetries
is usually less prone to experimental systematic errors like beam flux and
detector acceptance, high-accuracy data at moderate forward and backward
angles and energies $\lesssim120\;\MeV$ may provide good opportunities to
extract the nucleon polarisabilities. There, sensitivities are up to $\pm6\%$
for $\alphaes-\betams$, and so appreciable, asymmetries are large, and cross
sections (which determine rates) are still decent. Notice also that the
uncertainties from the \ChiEFT truncation and the sensitivities to the
polarisability combinations $\alphaes\pm\betams$ show different angular
dependence. On the other hand, potential- and polarisability-dependence show similar angular dependence. However, the sensitivity for $\alphaes-\betams=\pm2$ is about twice that for the potential. Therefore, forward angles can again be used to check data consistency with
the Baldin Sum Rule, and backward angles to extract $\alphaes-\betams$ at
potentially competitive levels. In that spirit, and without prejudicing
thorough experimental feasibility studies, the right-hand panels of figs.~\ref{fig:beamasymmetry}
to~\ref{fig:beamasym-pols} show the energy dependence of the beam asymmetry at two possible candidate angles,
$55^\circ$ and $110^\circ$ for the three canonical choices of $\alphaes\pm\betams$.
In these cases, the variation in $\Sigma^{\rm lin}$ from varying $\alphaes-\betams$ by $\pm2$ translates into rate-differences of $\pm1.6\;\mathrm{nb/sr}$.

\subsection{Comparison With Other Few-Nucleon Targets}
\label{sec:othertargets}

\begin{figure}[!b]
  \begin{center}
    \includegraphics[width=0.475\linewidth]
  {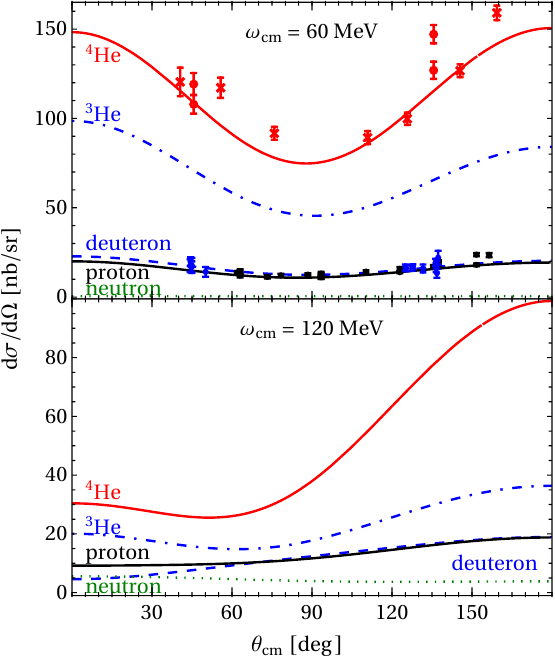}\hqq
    \includegraphics[width=0.485\linewidth]
  {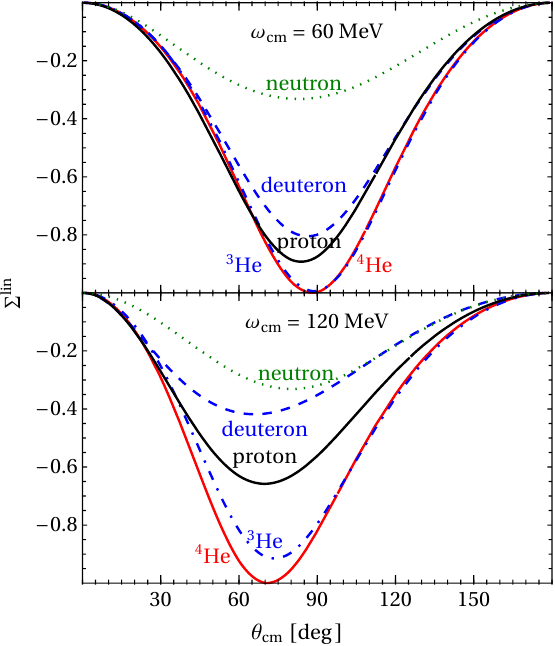}
  \caption{(Colour on-line) Comparison of the cross section (left) and beam
    asymmetry (right) in the centre-of-mass frame at $\omegacm=60\;\MeV$ (top)
    and $120\;\MeV$ (bottom) for a proton (black), neutron (green dotted),
    deuteron (blue dashed), \threeHe (blue dot-dashed) and \fourHe (black)
    target. The few-nucleon bound states use the same \NXLO{3}
    [$\calO(e^2\delta^3)$] kernels, including the same values for all
    polarisabilities, but slightly different $2\N$ and $3\N$ interactions. The
    cross section data at $60\;\MeV$ is taken on the proton (black) as
    discussed in~\cite{Griesshammer:2012we, McGovern:2012ew}, deuteron
    (blue)~\cite{Lundin:2002jy}, and \fourHe~(blue circle: \cite{WellsPhD};
    blue cross:~\cite{Sikora:2017rfk}) within $\pm5\;\MeV$ of the nominal
    energy and transformed into the centre-of-mass frame; see text for
    details.}
\label{fig:comparison}
\end{center}
\end{figure}

To put \fourHe Compton scattering in context, we contrast the predictions with
those for the proton, neutron, deuteron and \threeHe in
fig.~\ref{fig:comparison}. We will not reiterate the discussion of the latter
four targets in ref.~\cite{Margaryan:2018opu} but focus only on their comparison to
\fourHe. Each target is sensitive to different combinations of
nucleon polarisabilities: the proton and neutron, by definition, only to their particular
polarisabilities; the deuteron only to the isoscalar components
$\alphaep+\alphaen$ etc.~of both the scalar and spin polarisabilities;
\threeHe roughly to $2\alphaep+\alphaen$, $2\betamp+\betamn$ and to the spin
polarisabilities of the neutron but not of the proton; and \fourHe only to the
scalar-isoscalar combinations $\alphaep+\alphaen$ and $\betamp+\betamn$ but
not to any other. Each target thus provides access to complementary linear
combinations. 

Here, we focus on comparisons of the magnitudes of observables,
important for planning experiments.  At a minimum, they confirm that
data on the proton, deuteron and \fourHe are consistent with our unified theory description via the complete Compton kernel at
\NXLO{3} [$\calO(e^2\delta^3)$] and the same central values  of eq.~\eqref{eq:alphabeta} for the nucleon
polarisabilities. That the few-nucleon computations use slightly different $2\N$
and $3\N$ potentials, and the deuteron results include rescattering effects, is of small concern for the semi-quantitative
comparison we perform here in the the r\'egime $\omega\in[50;120]\;\MeV$. Likewise, there is little impact of the fact that the cross section data at $60\;\MeV$ have been
transformed into the centre-of-mass frame but not corrected for the difference
between actual and nominal experimental lab energy and angle, falling in a
corridor of $\pm5\;\MeV$ around the quoted energy. 

In contradistinction to the other plots, the
\emph{centre-of-mass (cm) frame} is used in fig.~\ref{fig:comparison}. This
avoids purely kinematic recoil effects which depend on target mass. For
example, the cross section that only includes the Thomson limit
[LO, $\calO(e^2\delta^0)$] is perfectly mirror-symmetric about $90^\circ$ in
the cm frame at all energies, while its forward angles are slightly higher than back-angles in
the lab frame, masking the actual amount of asymmetry caused by the target
and Compton physics. 

Figure~\ref{fig:comparison} clearly shows that Compton cross sections on
few-nucleon targets do not scale with powers of the target charge $Z$ in the
region $\omega\in[50;120]\;\MeV$ that is of interest for extracting
polarisabilities. While the proton and deuteron cross sections ($Z=1$) are of
roughly similar size, the \threeHe and \fourHe ones are vastly different
although they are both $Z=2$ targets. At $60\;\MeV$, the average of the \fourHe cross
section over $\cos\thetacm$ is about twice that of \threeHe and
nearly $7$ times of the proton and deuteron. At $120\;\MeV$, the ratio to
\threeHe is nearly unchanged while that to the proton and deuteron is about
$5$.  At these energies, the typical photon wave length $1/\omega$ is not
only comparable to the size of each nucleon constituent, but also to the range
$1/\mpi$ of the charged-pion exchange potential. Therefore, its effect is not ``frozen
out'' by such a high-energy probe and the photons interact
incoherently not only with each nucleon, but also with the charged long-range
part of the current which glues the nucleons into a coherent, bound nucleus.

Another striking feature is how asymmetric cross sections are. One simple
criterion is skewedness, which we define as the relative deviation of the cross section at
$180^\circ$ from that at $0^\circ$, measured as in eq.~\eqref{eq:reldev}. While at $60\;\MeV$, the cross section is, within $10\%$,
symmetric for all targets, the situation
is dramatically different at $\omegacm=120\;\MeV$: the huge skewedness of \fourHe (about $230\%$)
is only matched by that of the deuteron (about $300\%$), while both \threeHe
and the proton show only about $100\%$ relative deviation between forward and back
angles. The most likely reason is that the former are isoscalar
targets while the latter are $T=1/2$ ones. The rather large
isovector nucleon magnetic moment substantially reduces back-scattering,
countering the sizeable (isoscalar) dispersive correction from the magnetic
polarisability.
Angular skewedness is near-exclusively related to
the one-body amplitudes; the charged-meson-exchange currents show no
significant skewedness at any energy except in the much more loosely bound
deuteron.

We now consider the effect of varying the \emph{neutron} polarisability combination $\alphaen-\betamn$ by $\pm2$ units, in the \emph{cm frame}. 
For the deuteron and \fourHe 
this is exactly half of the variation of the
scalar-isoscalar polarisability combination $\alphaes-\betams$ by $\pm2$ units; \cf~sect.~\ref{sec:polarisabilities}. This
changes the backward-angle cross section in \fourHe at $120\;\MeV$ by
about $\mp5.5\%$ or $\gtrsim\pm5\;\mathrm{nb/sr}$. The same kinematics produces a variation of
about $\mp4\%$, \ie~$\lesssim\mp1.5\;\mathrm{nb/sr}$ in \threeHe and around $\gtrsim\mp3\%$
or less than $\mp0.6\;\mathrm{nb/sr}$ in the deuteron~\cite{Hildebrandt:2005ix, Hildebrandt:2005iw,
  Griesshammer:2013vga, Margaryan:2018opu}. So, heavier systems show both increased relative accuracy
and increased rates. 
The absolute variation for \fourHe and \threeHe is qualitatively quite different because,
while the polarisabilities interfere with the same
Thomson term of two protons, the charged-meson-exchange part is greater in a system with more possible $\n\p$ pairs. 
Likewise, that for the deuteron is
substantially smaller because the interference is with only one charged
constituent and charged-meson-exchange between one pair only. 
This change in nuclear effects like binding and meson-exchange currents across different targets allows for invaluable checks that these aspects of Compton scattering dynamics are well understood. 

The beam asymmetry $\Sigma^\mathrm{lin}$ is (except of course for the
neutron) dominated by the Thomson limit of scattering on a charged
point-particle at low energies. Figure~\ref{fig:comparison} shows that this
effect is most prominent for \fourHe and survives to higher energies than for
the other targets. The extremum  appears at roughly the same position for each
charged target, but only in \fourHe is the minimum value of about $-1$
practically unchanged from the Thomson limit over the energy range considered. At $120\;\MeV$, the less-tightly bound systems
only reach $-0.9$ (\threeHe) and $-0.4$ (deuteron), and the unbound proton has
$-0.65$. This may yet again demonstrate the competing effects of one-body and
charged-meson-exchange contributions. The angle-dependent magnitude of the sensitivity to varying the neutron's scalar polarisability combination $\alphaen-\betamn$ is not dissimilar in all cases, namely up to $\pm0.02$ in magnitude or up to $\pm3\%$ in relative deviation for \fourHe, about $3/4$ of that for \threeHe, and about half for the deuteron. Rate differences at $110^\circ$ are about $\pm0.8\;\mathrm{nb/sr}$ in\fourHe, with $\pm0.25\;\mathrm{nb/sr}$, i.e., less than a third of that in \threeHe, and at $\pm0.07\;\mathrm{nb/sr}$
in the deuteron.

A more comprehensive and rigorous discussion of target effects is left to an
upcoming cross-target study of observables and their sensitivities on the
polarisabilities~\cite{future}.

\section{Summary and Conclusions}
\label{sec:conclusions}

This presentation reports the first \emph{ab initio} calculation  (as defined in ref.~\cite{Ekstrom:2022yea}) of elastic Compton scattering on \fourHe. It is carried out in \ChiEFT at $\calO(e^2 \delta^3)$ [\NXLO{3}] in the $\delta$-expansion, as applied in the range $50\;\MeV \lesssim \omegalab \lesssim 120\;\MeV$. Since the scalar-isoscalar combinations of nucleon electric and magnetic polarisabilities that have already been measured in $\gamma$-deuteron scattering are exactly those that enter in this process, we take the deuteron values as input. Our calculation then has no free parameters. Compared to data from Illinois, MAXlab and \HIGS between $50$ and $80\;\MeV$, it predicts the correct angular dependence of the Compton differential cross section, and the right size. The energy dependence is correct within combined theory and experimental error bars, albeit the central values seem to have a different trend than the data. This overall agreement is a strong validation of \ChiEFT's ability to describe electromagnetic processes on light nuclei. 

We predict that Compton cross sections on the deuteron~\cite{Hildebrandt:2005ix,Hildebrandt:2005iw},
\threeHe~\cite{Margaryan:2018opu}, and \fourHe are approximately in the ratio $1:2:5$ at $\omega=120\;\MeV$. This scaling is not explained by the number of constituent charges. Rather, it provides concrete
evidence for the crucial r\^ole that pion-exchange currents, and, relatedly, $\n\p$ pairs correlated by one-pion exchange, play in determining the electromagnetic response in these nuclei for photons with energies of order $100\;\MeV$~\cite{Pastore:2019urn}. 

While this study may not suffice to reliably \emph{extract} polarisabilities from \fourHe data, it \emph{does} suffice for reliable rate estimates and, in particular, sensitivity studies. These may be useful for planning experiments. Observables are only sensitive to the two scalar-isoscalar dipole polarisabilities of the nucleon. For the cross section, we found that the relative sensitivity to variations of the less-well-known combination $\alphaes-\betams$ carries considerably smaller uncertainties than absolute rates. That is in part because residual theory errors are largely angle-independent, while the impact of this combination is
zero at forward angles and maximal at backward angles. The dependence of the cross section on a variation by $\pm2$ canonical units rises steadily from $\pm3\%$ of an overall $120\;\mathrm{nb}\,\mathrm{sr}^{-1}$ at $60\;\MeV$ 
to $\pm11\%$ of about $80\;\mathrm{nb}\,\mathrm{sr}^{-1}$ at $120\;\MeV$. 

Therefore, we advocate for data at high energies and large angles to facilitate high-accuracy extractions of the scalar-isoscalar polarisability combination $\alphaes-\betams$---this is also important for the deuteron and \threeHe. These should be complemented in the same experiment with measurements at low energies and small angles: They provide an important check of consistency with the Thomson limit and with the Baldin Sum Rule.

Measuring a ratio between forward- and back-angle cross sections may reduce theoretical uncertainties, while concurrently reducing angle-independent/systematic experimental errors.  We are cautious about the usefulness of beam-asymmetry data by itself. There, theory uncertainties and polarisability sensitivities show a very similar angular dependence, and rate-differences are quite small.

We are confident in these findings, based on an assessment of the region of validity and theoretical uncertainties of our approach. In the region in which the present formalism is applicable, $50\;\MeV\lesssim\omega\lesssim120\;\MeV$, we determined its accuracy to be about $\pm10\%$ across energies, with possibly $\pm12\%$ in back-scattering at the highest energies. The main theory uncertainty comes from the wave-function dependence across four \ChiEFT potentials with different cutoffs, 
and is of higher order.
While it is also angle-independent, we cannot use that fact to eliminate this source of theory uncertainties without additional theoretical justification.
Based on experience from the deuteron case, the inclusion of the mechanisms that restore the Thomson limit should also reduce the dependence on the choice of \ChiEFT interaction at the higher energies studied here. 

This is the other respect in which the calculation should be improved, namely by including coherent propagation of the $A$-body system between interactions with the first and second Compton photon incorporated into the calculation.
In the deuteron case, including these
effects that restore the nuclear Thomson limit leads to a reduction of the cross section by $10\%$
to $20\%$ at $\omega=50\;\MeV$, but only a few percent at $120\;\MeV$~\cite{Hildebrandt:2005ix, Hildebrandt:2005iw}.
For \fourHe, it is plausible that the corrections by coherent-nuclear effects
may suppress the cross section at the low end of our energy range somewhat
more: the mismatch between the Thomson-limit and the $\omega=0$ amplitude in
our calculation is larger than for either the deuteron or \threeHe, and
\fourHe has a substantially larger binding energy, so coherent propagation of
the four-nucleon system is likely to be important up to higher energies than
in the two- and three-nucleon case. We note that, in fact, our results already compare well with data even at $\omega \approx 50\;\MeV$.  Work to restore the Thomson limit in few-nucleon systems is in
progress~\cite{Walet:2023myk}. 

Our calculation of \fourHe Compton scattering  with full treatment of nuclear structure and two-body currents is tractable because we use the Transition Density Method that was developed and tested in refs.~\cite{Griesshammer:2020ufp, deVries:2023hin}. As was the case for \threeHe and \threeH, the 
one- and two-body densities generated for this investigation from $5$
chiral potentials as well as the AV18$+$UIX potential can be used for a
cornucopia of computations of reactions involving the \fourHe system. They are
available using a python package from  \url{https://pypi.org/project/nucdens/}. The method also opens the way to calculations of Compton scattering on other light nuclear targets like ${}^6$Li. 

The present investigation is part of the ongoing effort to develop a unified picture of Compton scattering on light nuclei~\cite{Beane:1999uq, Beane:2004ra,Hildebrandt:2004hh, Choudhury:2007bh, Shukla:2008zc, ShuklaPhD, Margaryan:2018opu}. 
The calculations carried out here use the same kernels as those of Compton scattering on \threeHe~\cite{Margaryan:2018opu}, but are one order lower than those that have been completed for the proton~\cite{McGovern:2012ew} and are in progress for the deuteron~\cite{itscomingreally}. Calculations at $\calO(e^2 \delta^4)$ are needed for a precision extraction of $\alphaes$ and $\betams$ that will have theory error bars concomitant with those arising from expected future high-quality \fourHe data. Such a calculation, and its comparison with forthcoming data~\cite{Hornidge, Feldman}, should permit a more straightforward extraction of $\alphaes - \betams$ than was possible in deuterium, because the effect of the nucleon spin polarisabilities is very small in a scalar-isoscalar nucleus like \fourHe. 

This work opens up the possibility of validating the \ChiEFT treatment 
of elastic Compton scattering across a range of few-nucleon systems. Such comparisons may also aid the understanding of both systematic errors in the measurements and the r\^ole of charged-pion-exchange currents across few-nucleon systems: from the loosely bound deuteron via \threeHe to tightly bound \fourHe. 
The deuteron theory is well-understood and sensitive to the same linear combination of polarisabilities as \fourHe~\cite{Griesshammer:2012we}.
\threeHe computations are now available as well, probing a different 
combination of neutron and proton polarisabilities, $2\alphaep+\alphaen$ \etc~\cite{Choudhury:2007bh,
  Shukla:2008zc, ShuklaPhD, Margaryan:2018opu}. Deuteron,
\fourHe and $^6$Li data is already available and new experiments are approved for these three targets and for 
\threeHe~\cite{Feldman, Hornidge}. Ultimately, a global fit across different nuclei would minimise the 
impact of experimental systematics in any one data set, exploit the different linear combinations of proton and neutron polarisabilities embedded in different nuclear targets, and test \ChiEFT's predictions for the impact of nuclear binding
and charged-meson-exchange currents---all while measuring a fundamental process that characterises the strong-interaction response of the constituents of nucleons and light nuclei to electromagnetic fields.

\section*{Author Contributions}

All authors shared equally in all tasks.

\section*{Data Availability Statement}

All data underlying this work are available
in full upon request from the corresponding author. The one- and two-body densities generated from $5$ chiral potentials and the AV18$+$UIX potential
 are available using the python package provided at \url{https://pypi.org/project/nucdens/}.

\section*{Code Availability Statement}

The one- and two-body densities generated from $5$ chiral potentials and the AV18$+$UIX potential
 are available using the python package provided at
 \url{https://pypi.org/project/nucdens/}, with explanations how to use
 them. Convolution codes underlying this work are available upon request from the corresponding author.

\section*{Declarations}

\section*{Competing Interests}

The authors declare no competing interests.

\newpage
\section*{Acknowledgements}

We are particularly indebted to our experimental colleagues M.~Ahmed, E.~J.~Downie, G.~Feldman and M.~H.~Sikora for discussions and encouragement. G.~Feldman and M.~H.\ Sikora tracked down the original sources of Compton data from the previous millennium. A.~Long provided comments on the manuscript.
We appreciate the warm hospitality and financial support for stays which were instrumental for this research: HWG at the University of Manchester, Ohio University and FZ J\"ulich; and DRP at George Washington University and Chalmers University of Technology. 
HWG is grateful to the organisers and participants of the meeting of MAMI's A2 collaboration in Mainz for the stimulating atmosphere and financial support. He also thanks the organisers and participants of MENU 2023 in Mainz for the opportunity to present preliminary results and for a delightful atmosphere.
This work was supported in part by the US Department of Energy under contract DE-SC0015393 (HWG, JL) and DE-FG02-93ER-40756 (DRP), by the UK Science and Technology Facilities Council grant ST/P004423/1 (JMcG), by the Deutsche Forschungsgemeinschaft and the Chinese National Natural Science Foundation through funds provided to the Sino-German CRC 110 ``Symmetries and the Emergence of Structure in QCD'' (AN; DFG grant TRR~110; NSFC grant 11621131001), by the Ministerium f\"ur Kultur und Wissenschaft Nordrhein-Westphalen (MKW-NW) under funding code NW21-024-A (AN) and by a Tage Erlander Professorship from the Swedish Research Council, grant 2022-00215 (DRP). Additional funds for HWG were provided by an award of the High Intensity Gamma-Ray Source \HIGS of the Triangle Universities Nuclear Laboratory TUNL in concert with the Department of Physics of Duke University, and by George Washington University: by the Office of the Vice President for Research and the Dean of the Columbian College of Arts and Sciences; by an Enhanced Faculty Travel Award of the Columbian College of Arts and Sciences. His research was conducted in part in GW's Campus in the Closet.
The computations of nuclear densities were performed on \textsc{Jureca} of the J\"ulich Supercomputing Centre (J\"ulich, Germany).


\appendix

\begin{thebibliography}{99}



\bibitem{Howell:2020nob}
  C.~R.~Howell, M.~W.~Ahmed, A.~Afanasev, D.~Alesini, J.~R.~M.~Annand, A.~Aprahamian, D.~L.~Balabanski, S.~V.~Benson, A.~Bernstein and C.~R.~Brune, \textit{et al.}
  J. Phys. G \textbf{49} (2022) 010502
  \doi{10.1088/1361-6471/ac2827}
  \arXiv[arXiv:2012.10843 [nucl-ex]].

\bibitem{Griesshammer:2012we} H.~W.~Grie{\ss}hammer, J.~A.~McGovern,
  D.~R.~Phillips and G.~Feldman,
  Prog.\ Part.\ Nucl.\ Phys.\ {\bf 67} (2012) 841
  \doi{10.1016/j.ppnp.2012.04.003}
  \arXiv[arXiv:1203.6834 [nucl-th]].

\bibitem{Myers:2014ace} L.~S.~Myers {\it et al.}  [COMPTON@MAX-lab
  Collaboration],
  Phys.\ Rev.\ Lett.\ {\bf 113} (2014) 262506 
  \doi{10.1103/PhysRevLett.113.262506}
  \arXiv[arXiv:1409.3705 [nucl-ex]].
  

\bibitem{Myers:2015aba} L.~S.~Myers {\it et al.},
  Phys.\ Rev.\ C {\bf 92} (2015) 025203
  \doi{10.1103/PhysRevC.92.025203}
  \arXiv[arXiv:1503.08094 [nucl-ex]].
  

\bibitem{Griesshammer:2015ahu} H.~W.~Grie\3hammer, J.~A.~McGovern and
  D.~R.~Phillips,
  Eur.\ Phys.\ J.\ A \textbf{52} (2016) 139 
  \doi{10.1140/epja/i2016-16139-5} \doi{10.22323/1.253.0104}
  \arXiv[arXiv:1511.01952 [nucl-th]].

\bibitem{OlmosdeLeon:2001zn}
  V.~Olmos de Leon {\it et al.},
  Eur.\ Phys.\ J.\ A {\bf 10} (2001) 207
  \doi{10.1007/s100500170132}.

\bibitem{Levchuk:1999zy} M.~I.~Levchuk and A.~I.~L'vov,
  Nucl.\ Phys.\ A {\bf 674} (2000) 449 
  \doi{10.1016/S0375-9474(00)00145-7}
  \arXivold[nucl-th/9909066].
  

\bibitem{Gryniuk:2015eza} 
  O.~Gryniuk, F.~Hagelstein and V.~Pascalutsa,
  Phys.\ Rev.\ D {\bf 92} (2015) 074031
  \doi{10.1103/PhysRevD.92.074031}
  \arxiv[arXiv:1508.07952 [nucl-th]].

\bibitem{Detmold:2019ghl} 
  W.~Detmold {\it et al.} [USQCD Collaboration],
  Eur.\ Phys.\ J.\ A {\bf 55} (2019) 193
  \doi{10.1140/epja/i2019-12902-4}
  \arXiv[arXiv:1904.09512 [hep-lat]].

\bibitem{Alexandru:2019} 
  A.~Alexandru,
  PoS CD {\bf 2018} (2019) 021
  \doi{10.22323/1.317.0021}.

\bibitem{Bignell:2020xkf} 
  R.~Bignell, W.~Kamleh and D.~Leinweber,
  ``Magnetic polarizability of the nucleon using a Laplacian mode projection,''
  Phys.\ Rev.\ D {\bf 101} (2020) 094502
  \doi{10.1103/PhysRevD.101.094502}
  \arXiv[arXiv:2002.07915 [hep-lat]].

\bibitem{Wilcox:2021rtt}
    W.~Wilcox and F.~X.~Lee,
    Phys. Rev. D \textbf{104} (2021) 034506
    \doi{10.1103/PhysRevD.104.034506}
    \arXiv[arXiv:2106.02557 [hep-lat]].

\bibitem{Wang:2023omf}
    X.~H.~Wang, C.~L.~Fan, X.~Feng, L.~C.~Jin and Z.~L.~Zhang,
    \arXiv[arXiv:2310.01168 [hep-lat]].

\bibitem{Gasser:2015dwa} 
  J.~Gasser, M.~Hoferichter, H.~Leutwyler and A.~Rusetsky,
  Eur.\ Phys.\ J.\ C {\bf 75} (2015) 375
  \doi{10.1140/epjc/s10052-015-3580-9}
  \arXiv[arXiv:1506.06747 [hep-ph]].

\bibitem{Thomas:2014dxa} 
  A.~W.~Thomas, X.~G.~Wang and R.~D.~Young,
  Phys.\ Rev.\ C {\bf 91} (2015) 015209
  \arxiv[arXiv:1406.4579 [nucl-th]].

\bibitem{Tomalak:2018dho} 
  O.~Tomalak,
  Eur.\ Phys.\ J.\ Plus {\bf 135} (2020) 411
  \doi{10.1140/epjp/s13360-020-00413-9}
  \arXiv[arXiv:1810.02502 [hep-ph]].

\bibitem{Walker-Loud:2019qhh} 
  A.~Walker-Loud,
  ``On the Cottingham formula and the electromagnetic contribution to the proton-neutron mass splitting,''
  PoS CD {\bf 2018} (2019) 045
  \doi{10.22323/1.317.0045}
  \arXiv[arXiv:1907.05459 [nucl-th]].

\bibitem{Gasser:2020mzy} 
  J.~Gasser, H.~Leutwyler and A.~Rusetsky,
  ``On the mass difference between proton and neutron,''
  Phys.\ Lett.\ B {\bf 814} (2021) 136087
  \doi{10.1016/j.physletb.2021.136087}
  \arXiv[arXiv:2003.13612 [hep-ph]].
  

\bibitem{Mornacchi:2022cln}
    E.~Mornacchi, S.~Rodini, B.~Pasquini and P.~Pedroni,
    Phys. Rev. Lett. \textbf{129} (2022) 102501
    \doi{10.1103/PhysRevLett.129.102501}
    \arXiv[arXiv:2204.13491 [hep-ph]].

\bibitem{Ekstrom:2022yea}
    A.~Ekstr\"om, C.~Forss\'en, G.~Hagen, G.~R.~Jansen, W.~Jiang and T.~Papenbrock,
    Front. Phys. \textbf{11} (2023), 1129094
    \doi{10.3389/fphy.2023.1129094}
    \arXiv[arXiv:2212.11064 [nucl-th]].

\bibitem{Sikora:2017rfk}
  M.~H.~Sikora, M.~W.~Ahmed, A.~Banu, C.~Bartram, B.~Crowe, E.~J.~Downie, G.~Feldman, H.~Gao, H.~W.~Grie\3hammer and H.~Hao, \textit{et al.}
  Phys. Rev. C \textbf{96} (2017) 055209
  \doi{10.1103/PhysRevC.96.055209}

\bibitem{Sikora}
    M.~H.~Sikora, private communication (2020).

\bibitem{Griesshammer:2020ufp} 
  H.~W.~Grie\3hammer, J.~A.~McGovern, A.~Nogga and D.~R.~Phillips:
  Few-Body Syst.~\textbf{~61} (2020) 48
  \doi{10.1007/s00601-020-01578-w}
  \arXiv[arXiv:2005.12207 [nucl-th]].

\bibitem{deVries:2023hin}
J.~de Vries, C.~K\"ober, A.~Nogga and S.~Shain,
\arXiv[arXiv:2310.11343 [hep-ph]].

\bibitem{Margaryan:2018opu} 
  A.~Margaryan, B.~Strandberg, H.~W.~Grie\3hammer, J.~A.~Mcgovern, D.~R.~Phillips and D.~Shukla,
  Eur.\ Phys.\ J.\ A {\bf 54} (2018) 125
  \doi{10.1140/epja/i2018-12554-x}
  \arXiv[arXiv:1804.00956 [nucl-th]].

\bibitem{Beane:1999uq} 
  S.~R.~Beane, M.~Malheiro, D.~R.~Phillips and U.~van Kolck,
  Nucl.\ Phys.\ A {\bf 656} (1999) 367
  \doi{10.1016/S0375-9474(99)00312-7}
  \arXivold[nucl-th/9905023].

\bibitem{Choudhury:2007bh} D.~Choudhury, A.~Nogga and D.~R.~Phillips,
  Phys.\ Rev.\ Lett.\ {\bf 98} (2007) 232303
  \doi{10.1103/PhysRevLett.98.232303}
  \arXivold[nucl-th/0701078].

\bibitem{Shukla:2008zc} 
  D.~Shukla, A.~Nogga and D.~R.~Phillips,
  Nucl.\ Phys.\ A {\bf 819} (2009) 98
  \doi{10.1016/j.nuclphysa.2009.01.003}
  \arXiv[arXiv:0812.0138 [nucl-th]].

\bibitem{ShuklaPhD} D.~Choudhury, \emph{PhD thesis}, Ohio University (2006)
  \href{http://rave.ohiolink.edu/etdc/view?acc_num=ohiou1163711618}
  {http://rave.ohiolink.edu/etdc/\disc{}view?acc\_num=ohiou1163711618}.

\bibitem{MENU2023}
  H.~W.~Grie\3hammer, J.~Liao, J.~A.~McGovern, A.~Nogga and D.~R.~Phillips, Proceedings of the 16th International Conference on Meson-Nucleon Physics and the Structure of the Nucleon (MENU 2023), submitted \arXiv[arXiv:2401.15673 [nucl-th]].
  
\bibitem{Long}
  A.~Long, H.~W.~Grie\3hammer and A.~Nogga, in progress.

\bibitem{Shukla:2018rzp} 
  D.~Choudhury, A.~Nogga and D.~R.~Phillips,
  Phys.\ Rev.\ Lett.\  {\bf 98} (2007) 232303
  Erratum: [Phys.\ Rev.\ Lett.\  {\bf 120} (2018) 249901]
  \doi{10.1103/PhysRevLett.120.249901}, \doi{10.1103/PhysRevLett.98.232303}
  \arXiv[arXiv:1804.01206 [nucl-th]], \arXivold[nucl-th/0701078]].

\bibitem{Weinberg:1990rz} 
  S.~Weinberg,
  Phys.\ Lett.\ B {\bf 251} (1990) 288
  \doi{10.1016/0370-2693(90)90938-3}.

\bibitem{Weinberg:1991um}
  S.~Weinberg,
  Nucl. Phys. B \textbf{363} (1991) 3
  \doi{10.1016/0550-3213(91)90231-L}.

\bibitem{Weinberg:1992yk}
  S.~Weinberg,
  Phys. Lett. B \textbf{295} (1992) 114
  \doi{10.1016/0370-2693(92)90099-P}
  \arXivold[arXiv:hep-ph/9209257 [hep-ph]].
  

\bibitem{vanKolckPhD} U.~van Kolck, \emph{PhD thesis}, University of Texas at
  Austin (1993). 

\bibitem{vanKolck:1994yi} 
  U.~van Kolck,
  Phys.\ Rev.\ C {\bf 49} (1994) 2932
  \doi{10.1103/PhysRevC.49.2932}.

\bibitem{Friar:1996zw} 
  J.~L.~Friar,
  Few Body Syst.\  {\bf 22} (1997) 161
  \doi{10.1007/s006010050059}
  \arXivold[nucl-th/9607020].

\bibitem{Epelbaum:2008ga} 
  E.~Epelbaum, H.-W.~Hammer and U.~G.~Mei\3ner,
  Rev.\ Mod.\ Phys.\  {\bf 81} (2009) 1773
  \doi{10.1103/RevModPhys.81.1773}
  \arXiv[arXiv:0811.1338 [nucl-th]].

\bibitem{Phillips:2016mov}
  D.~R.~Phillips,
  Ann. Rev. Nucl. Part. Sci. \textbf{66} (2016) 421
  \doi{10.1146/annurev-nucl-102014-022321}.
  

\bibitem{Machleidt:2016rvv} 
  R.~Machleidt and F.~Sammarruca,
  Phys.\ Scripta {\bf 91} (2016) 083007
  \doi{10.1088/0031-8949/91/8/083007}
  \arXiv[arXiv:1608.05978 [nucl-th]].

\bibitem{Hammer:2019poc} 
  H.-W.~Hammer, S.~K\"onig and U.~van Kolck,
  Rev.\ Mod.\ Phys.\  {\bf 92} (2020) 025004
  \doi{10.1103/RevModPhys.92.025004}
  \arXiv[arXiv:1906.12122 [nucl-th]].

\bibitem{Epelbaum:2019kcf} 
  E.~Epelbaum, H.~Krebs and P.~Reinert,
  Front.\ in Phys.\  {\bf 8} (2020) 98
  \doi{10.3389/fphy.2020.00098}
  \arXiv[arXiv:1911.11875 [nucl-th]].
  

\bibitem{Edmonds} A.~R.~Edmonds, ``Angular Momentum in Quantum Mechanics'',
  Princeton University Press (1974).

\bibitem{PDG}
  R.~L.~Workman \textit{et al.} [Particle Data Group],
  PTEP \textbf{2022} (2022) 083C01
  \doi{10.1093/ptep/ptac097} and \url{http://pdg.lbl.gov}.

\bibitem{Polyzou:1997je}
    W.~N.~Polyzou,
    Phys. Rev. C \textbf{58} (1998), 91-95
    \doi{10.1103/PhysRevC.58.91}
    \arXiv[arXiv:nucl-th/9711046 [nucl-th]].

\bibitem{More:2017syr}
    S.~N.~More, S.~K.~Bogner and R.~J.~Furnstahl,
    Phys. Rev. C \textbf{96} (2017) 054004
    \doi{10.1103/PhysRevC.96.054004}
    \arXiv[arXiv:1708.03315 [nucl-th]].

\bibitem{Tropiano:2021qgf}
    A.~J.~Tropiano, S.~K.~Bogner and R.~J.~Furnstahl,
    Phys. Rev. C \textbf{104} (2021) 034311
    \doi{10.1103/PhysRevC.104.034311}
    \arXiv[arXiv:2105.13936 [nucl-th]].



\bibitem{McGovern:2012ew}
  J.~A.~McGovern, D.~R.~Phillips and H.~W.~Grie\3hammer,
  Eur. Phys. J. A \textbf{49} (2013) 12
  \doi{10.1140/epja/i2013-13012-1}
  \arXiv[arXiv:1210.4104 [nucl-th]].

\bibitem{Hildebrandt:2005ix}
  R.~P.~Hildebrandt, \emph{PhD thesis}, Technische Universit\"at M\"unchen (2005)
 \arxivold[nucl-th/0512064].

\bibitem{Hildebrandt:2005iw} R.~P.~Hildebrandt, H.~W.~Grie\3hammer and
  T.~R.~Hemmert,
  Eur.\ Phys.\ J.\ A {\bf 46} (2010) 111 
  \doi{10.1140/epja/i2010-11024-y}
  \arXivold[nucl-th/0512063].

\bibitem{Furnstahl:2014xsa}
    R.~J.~Furnstahl, D.~R.~Phillips and S.~Wesolowski,
    J. Phys. G \textbf{42} (2015) no.3, 034028
    \doi{10.1088/0954-3899/42/3/034028}
    \arXiv[arXiv:1407.0657 [nucl-th]].

\bibitem{Griesshammer:2021zzz}
  H.~W.~Grie\3hammer,
  Few Body Syst. \textbf{63} (2022) 44
  \doi{10.1007/s00601-022-01739-z}
  \arXiv[arXiv:2111.00930 [nucl-th]].

\bibitem{Pascalutsa:2002pi}
  V.~Pascalutsa and D.~R.~Phillips,
  Phys.\ Rev.\  C {\bf 67} (2003) 055202
  \arxivold[arXiv:nucl-th/0212024].
  

\bibitem{Beane:2004ra}
  S.~R.~Beane, M.~Malheiro, J.~A.~McGovern, D.~R.~Phillips and U.~van Kolck,
  Nucl.\ Phys.\  A {\bf 747} (2005) 311
  \doi{10.1016/j.nuclphysa.2004.09.068}
  \arXivold[arXiv:nucl-th/0403088].

\bibitem{Bernard:1991rq} V.~Bernard, N.~Kaiser, U.~G.~Mei{\ss}ner,
  Phys.\ Rev.\ Lett.\ {\bf 67 } (1991) 1515
  \doi{10.1103/PhysRevLett.67.1515}.
  

\bibitem{Bernard:1995dp} V.~Bernard, N.~Kaiser and U.~G.~Mei{\ss}ner,
  Int.\ J.\ Mod.\ Phys.\ E {\bf 4} (1995) 193 
  \doi{10.1142/S0218301395000092}
  \arXivold[arXiv:hep-ph/9501384].

\bibitem{Hildebrandt:2003fm} R.~P.~Hildebrandt, H.~W.~Grie{\ss}hammer,
  T.~R.~Hemmert and B.~Pasquini,
  Eur.\ Phys.\ J.\ A {\bf 20} (2004) 293 
  \doi{10.1140/epja/i2003-10144-9}
  \arXivold[arXiv:nucl-th/0307070].

\bibitem{Hildebrandt:2004hh} R.~P.~Hildebrandt, H.~W.~Grie\3hammer,
  T.~R.~Hemmert and D.~R.~Phillips,
  Nucl.\ Phys.\ A {\bf 748} (2005) 573
  \doi{10.1016/j.nuclphysa.2004.11.017}
  \arXivold[nucl-th/0405077].
  

\bibitem{Griesshammer:2013vga}
  H.~W.~Grie\3hammer,
  Eur. Phys. J. A \textbf{49} (2013) 100
  [erratum: Eur. Phys. J. A \textbf{53} (2017) 113; erratum: Eur. Phys. J. A \textbf{54} (2018) 57]
  \doi{10.1140/epja/i2013-13100-2}, \doi{10.1140/epja/i2017-12311-9}, \doi{10.1140/epja/i2018-12502-x}
  \arXiv[arXiv:1304.6594 [nucl-th]].

\bibitem{Butler:1992ci}
  M.~N.~Butler and M.~J.~Savage,
  Phys.\ Lett.\ B {\bf 294} (1992) 369
  \doi{10.1016/0370-2693(92)91535-H}
  \arXivold[hep-ph/9209204].

\bibitem{Hemmert:1996rw}
  T.~R.~Hemmert, B.~R.~Holstein and J.~Kambor,
  Phys.\ Rev.\ D {\bf 55} (1997) 5598 
  \doi{10.1103/PhysRevD.55.5598}
  \arXivold[hep-ph/9612374].

\bibitem{Hemmert:1997tj}
  T.~R.~Hemmert, B.~R.~Holstein, J.~Kambor and G.~Knochlein,
  Phys.\ Rev.\ D {\bf 57} (1998) 5746 
  \doi{10.1103/PhysRevD.57.5746}
  \arXivold[nucl-th/9709063].

\bibitem{Griesshammer:2017txw} H.~W.~Grie\3hammer, J.~A.~McGovern and
  D.~R.~Phillips,
  Eur.\ Phys.\ J.\ A {\bf 54} (2018) 37
  \doi{10.1140/epja/i2018-12467-8}
  \arXiv[arXiv:1711.11546 [nucl-th]]. 

\bibitem{Reinert:2017usi} 
  P.~Reinert, H.~Krebs and E.~Epelbaum,
  Eur.\ Phys.\ J.\ A {\bf 54} (2018) 86
  \doi{10.1140/epja/i2018-12516-4}
  \arXiv[arXiv:1711.08821 [nucl-th]].

\bibitem{Maris:2020qne}
 P.~Maris, E.~Epelbaum, R.~J.~Furnstahl, J.~Golak, K.~Hebeler, T.~H\"uther, H.~Kamada,  H.~Krebs, U.~G.~Mei\ss{}ner, J.~A.~Melendez, \textit{et al.}
 Phys. Rev. C \textbf{103} (2021) 054001
 \doi{10.1103/PhysRevC.103.054001}
 \arXiv[arXiv:2012.12396 [nucl-th]].

\bibitem{Le:2023bfj}
 H.~Le, J.~Haidenbauer, U.~G.~Mei\ss{}ner and A.~Nogga,
 Eur. Phys. J. A  \textbf{60} (2024) 3
 \doi{10.1140/epja/s10050-023-01219-w}
 \arXiv[arXiv:2308.01756 [nucl-th]].

\bibitem{Phillips:2013fia} 
  D.~R.~Phillips,
  PoS CD {\bf 12} (2013) 013
  \doi{10.22323/1.172.0013}
  \arxiv[arXiv:1302.5959 [nucl-th]].

\bibitem{vanKolck:2020llt}
    U.~van Kolck,
    Front. in Phys. \textbf{8} (2020) 79
    \doi{10.3389/fphy.2020.00079}
    \arXiv[arXiv:2003.06721 [nucl-th]].

\bibitem{Tews:2022yfb}
    I.~Tews, Z.~Davoudi, A.~Ekstr\"om, J.~D.~Holt, K.~Becker, R.~Brice\~no, D.~J.~Dean, W.~Detmold, C.~Drischler and T.~Duguet, \textit{et al.}
    Few Body Syst. \textbf{63} (2022) 67
    \doi{10.1007/s00601-022-01749-x}
    \arXiv[arXiv:2202.01105 [nucl-th]].

\bibitem{Wiringa:1994wb} 
  R.~B.~Wiringa, V.~G.~J.~Stoks and R.~Schiavilla,
  Phys.\ Rev.\ C {\bf 51} (1995) 38
  \doi{10.1103/PhysRevC.51.38}
  \arXivold[nucl-th/9408016].

\bibitem{Pudliner:1995wk} 
  B.~S.~Pudliner, V.~R.~Pandharipande, J.~Carlson and R.~B.~Wiringa,
  Phys.\ Rev.\ Lett.\  {\bf 74} (1995) 4396
  \doi{10.1103/PhysRevLett.74.4396}
  \arXivold[nucl-th/9502031].

\bibitem{Pudliner:1997ck} 
  B.~S.~Pudliner, V.~R.~Pandharipande, J.~Carlson, S.~C.~Pieper and R.~B.~Wiringa,
  Phys.\ Rev.\ C {\bf 56} (1997) 1720
  \doi{10.1103/PhysRevC.56.1720}
  \arXivold[nucl-th/9705009].

\bibitem{Entem:2003ft} 
  D.~R.~Entem and R.~Machleidt,
  Phys.\ Rev.\ C {\bf 68} (2003) 041001
  \doi{10.1103/PhysRevC.68.041001}
  \arXivold[nucl-th/0304018].

\bibitem{Nogga:2005hp} 
  A.~Nogga, P.~Navratil, B.~R.~Barrett and J.~P.~Vary,
  Phys.\ Rev.\ C {\bf 73} (2006) 064002
  \doi{10.1103/PhysRevC.73.064002}
  \arXivold[nucl-th/0511082].

\bibitem{WellsPhD}
  D.~P.~Wells, 
  \emph{PhD thesis}, University of Illinois at Urbana-Champaign (1990).

\bibitem{Fuhrberg:1995zz} 
  K.~Fuhrberg \etal,
  Nucl.\ Phys.\ A {\bf 591} (1995) 1
  \doi{10.1016/0375-9474(95)00177-3}.

\bibitem{Li:2019irp} 
  X.~Li \etal, 
  Phys.\ Rev.\ C {\bf 101} (2020) 034618
  \doi{10.1103/PhysRevC.101.034618}
  \arXiv[arXiv:1912.06915 [nucl-ex]].


\bibitem{Furnstahl:2015rha}
  R.~J.~Furnstahl, N.~Klco, D.~R.~Phillips and S.~Wesolowski,
  Phys. Rev. C \textbf{92} (2015) 024005
  \doi{10.1103/PhysRevC.92.024005}
  \arXiv[arXiv:1506.01343 [nucl-th]].

\bibitem{Melendez:2019izc}
  J.~A.~Melendez, R.~J.~Furnstahl, D.~R.~Phillips, M.~T.~Pratola and S.~Wesolowski,
  Phys. Rev. C \textbf{100} (2019) 044001
  \doi{10.1103/PhysRevC.100.044001}
  \arXiv[arXiv:1904.10581 [nucl-th]].

\bibitem{Melendez:2020ikd}
  J.~A.~Melendez, R.~J.~Furnstahl, H.~W.~Grie\3hammer, J.~A.~McGovern, D.~R.~Phillips and M.~T.~Pratola,
    Eur.\ Phys.\ J.\ A \textbf{57} (2021) 81
  \doi{10.1140/epja/s10050-021-00382-2}
  \arXiv[arXiv:2004.11307 [nucl-th]].

\bibitem{Blanpied:1996}
  G.~Blanpied et al. [LEGS collaboration],
  Phys.\ Rev.\ Lett.\  {\bf 76} (1996) 1023
  \doi{10.1103/PhysRevLett.76.1023}.

\bibitem{Blanpied:2001}
  G.~Blanpied et al. [LEGS collaboration],
  Phys.\ Rev.\  {\bf C64} (2001) 025203
  \doi{10.1103/PhysRevC.64.025203}.

\bibitem{Martel-PhD:2013}
  P. P. Martel, \emph{PhD thesis}, University of Massachusetts Amherst (2013)  \doi{doi.org/10.7275/j1yn-de26}.

\bibitem{Sokhoyan:2016yrc} 
  V.~Sokhoyan {\it et al.} [A2 Collaboration],
  Eur.\ Phys.\ J.\ A {\bf 53} (2017) 14
  \doi{10.1140/epja/i2017-12203-0}
  \arxiv[arXiv:1611.03769 [nucl-ex]].

\bibitem{Collicott:2015} C.~Collicott, \emph{PhD thesis}, Dalhousie University
  (2015) \url{https://wwwa2.kph.uni-mainz.de/images/publications/phd/thesis_Collicott-Cristina-2015.pdf} .

\bibitem{Martel:2017pln} 
  P.~Martel {\it et al.} [A2 Collaboration],
  EPJ Web Conf.\  {\bf 142} (2017) 01021
  \doi{10.1051/epjconf/201714201021}.

\bibitem{HIGS2019} X.~Li, \etal,
  Phys. Rev. Lett. \textbf{128} (2022) 132502 
  \doi{10.1103/PhysRevLett.128.132502}
  \arXiv[arXiv:2205.10533 [nucl-ex]].

\bibitem{A2:2019bqm}
  D.~Paudyal \textit{et al.} [A2],
  Phys. Rev. C \textbf{102} (2020) no.3, 035205
  \doi{10.1103/PhysRevC.102.035205} 
  \arXiv[arXiv:1909.02032 [nucl-ex]].

\bibitem{A2CollaborationatMAMI:2021vfy}
  E.~Mornacchi \textit{et al.} [A2 Collaboration at MAMI],
  Phys. Rev. Lett. \textbf{128} (2022) no.13, 132503
  \doi{10.1103/PhysRevLett.128.132503} 
  \arXiv[arXiv:2110.15691 [nucl-ex]].

\bibitem{Jackson} J.D. Jackson, \emph{Classical Electrodynamics}, Wiley, 1998.

\bibitem{Lundin:2002jy}
  M.~Lundin, J.~O.~Adler, M.~Boland, K.~Fissum, T.~Glebe, K.~Hansen, L.~Isaksson, O.~Kaltschmidt, M.~Karlsson and K.~Kossert, \textit{et al.}
  Phys. Rev. Lett. \textbf{90} (2003) 192501
  \doi{10.1103/PhysRevLett.90.192501}
  \arXivold[arXiv:nucl-ex/0204014 [nucl-ex]].

\bibitem{future} 
  H.~W.~Grie\3hammer, J.~A.~McGovern, A.~Nogga and D.~R.~Phillips, in preparation.


\bibitem{Pastore:2019urn}
    S.~Pastore, J.~Carlson, S.~Gandolfi, R.~Schiavilla and R.~B.~Wiringa,
    Phys. Rev. C \textbf{101} (2020) 044612
    \doi{10.1103/PhysRevC.101.044612}
    \arXiv[arXiv:1909.06400 [nucl-th]].

\bibitem{itscomingreally}
  H.~W.~Grie\3hammer, J.~A.~McGovern and D.~R.~Phillips, in preparation.
  
  \bibitem{Hornidge} D.~Hornidge, Proceedings of the 16th International Conference on Meson-Nucleon Physics and the Structure of the Nucleon (MENU 2023), submitted.
  
  \bibitem{Feldman} G.~Feldman, Proceedings of the 16th International Conference on Meson-Nucleon Physics and the Structure of the Nucleon (MENU 2023), submitted.
  
\bibitem{Walet:2023myk}
  N.~R.~Walet, J.~Singh, J.~Kirscher, M.~C.~Birse, H.~W.~Grie\3hammer and
  J.~A.~McGovern,
    Few-Body Syst.~\textbf{64} (2023) 56
  \doi{10.1007/s00601-023-01824-x}
  \arXiv[arXiv:2303.09361 [nucl-th]].




\end{thebibliography}
\end{document}